\documentclass[11pt]{article}
\usepackage{jcapmod}

\usepackage{booktabs}
\usepackage[english]{babel}
\usepackage{amsmath,amssymb,amsbsy,amstext, amsthm, simplewick}
\usepackage{hyperref}
\usepackage{graphicx}
\usepackage{amsfonts}
\usepackage{amssymb}
\usepackage{upgreek}
\usepackage{simplewick}
 \usepackage{exscale,relsize}

\usepackage[margin=1cm,labelfont={sf,bf,scriptsize},textfont={sl,scriptsize}]{caption}

\usepackage{colortbl}
\definecolor{lightgreen}{cmyk}{0.2, 0, 0.2, 0.2}
\definecolor{lightgray}{cmyk}{0.1,0.2,0,0.1}
\definecolor{lightgray2}{cmyk}{0.1,0.1,0,0.1}

\makeatletter
\newlength{\apb@width}
\newcommand{\autoparbox}[2][c]{\settowidth{\apb@width}{#2}\parbox[#1]{\apb@width}{#2}}

\makeatother

\numberwithin{equation}{section}

\def\beq{\begin{equation}}
\def\eeq{\end{equation}}

\def\bea{\begin{eqnarray}}
\def\eea{\end{eqnarray}}

\def\d{{\rm d}}

\newcommand\lsim{\mathrel{\rlap{\lower4pt\hbox{\hskip1pt$\sim$}}
        \raise1pt\hbox{$<$}}}
\newcommand\gsim{\mathrel{\rlap{\lower4pt\hbox{\hskip1pt$\sim$}}
        \raise1pt\hbox{$>$}}}

\def\beq{\begin{equation}}
\def\eeq{\end{equation}}
\def\bea{\begin{eqnarray}}
\def\eea{\end{eqnarray}}

\def\d{{\rm d}}

\def\d{{\rm d}}

\def\h{{\rm h}}
\def\m{{\rm m}}

\def\k{{\boldsymbol{k}}}
\def\q{{\boldsymbol{q}}}

\def\x{{\boldsymbol{x}}}
\def\r{{\boldsymbol{r}}}

\def\M{\mathsmaller{M}}
\DeclareRobustCommand{\SkipTocEntry}[4]{}

\def\fnl{f_{\mathsmaller{\rm NL}}}
\def\gnl{g_{\mathsmaller{\rm NL}}}
\def\tnl{\tau_{\mathsmaller{ \rm NL}}}

\begin{document}

\begin{titlepage}

\setcounter{page}{1} \baselineskip=15.5pt \thispagestyle{empty}

\bigskip\

\vspace{2cm}
\begin{center}
{\fontsize{16}{28}\selectfont  \bf Stochastic Bias from Non-Gaussian Initial Conditions}
\end{center}

\vspace{0.2cm}

\begin{center}
{\fontsize{13}{30}\selectfont   Daniel Baumann$^{\bigstar}$, Simone Ferraro$^{\diamondsuit}$, Daniel Green$^{\clubsuit, \blacklozenge, \spadesuit}$, and Kendrick M.~Smith$^{\diamondsuit,\heartsuit}$}
\end{center}

%\vspace{0.2cm}

\begin{center}
\vskip 8pt
\textsl{$^\bigstar$ D.A.M.T.P., Cambridge University, Cambridge, CB3 0WA, UK}

\vskip 7pt
\textsl{$^\diamondsuit$ Princeton University Observatory, Peyton Hall, Ivy Lane, Princeton, NJ 08544, USA}

\vskip 7pt
\textsl{$^\clubsuit$ School of Natural Sciences,
 Institute for Advanced Study,
Princeton, NJ 08540, USA}

\vskip 7pt
\textsl{$^ \blacklozenge$
Stanford Institute for Theoretical Physics, Stanford University, Stanford, CA 94306, USA}

\vskip 7pt
\textsl{$^\spadesuit$ Kavli Institute for Particle Astrophysics and Cosmology, Stanford, CA 94025, USA}

\vskip 7pt
\textsl{$^\heartsuit$ Perimeter Institute for Theoretical Physics, Waterloo, ON N2L 2Y5, Canada}

\end{center}

\vspace{1.2cm}
\hrule \vspace{0.3cm}
{ \noindent \textbf{Abstract} \\[0.2cm]
\noindent 
In this article, we show that a stochastic form of scale-dependent halo bias arises in multi-source inflationary models, where multiple fields determine the initial curvature perturbation.
We derive this effect for general non-Gaussian initial conditions and  study various examples, such as curvaton models and quasi-single field inflation. 
We present a general formula for both the stochastic and the non-stochastic parts of the halo bias, in terms of the $N$-point cumulants of the curvature
perturbation at the end of inflation.  At lowest order, the stochasticity arises if the collapsed limit of the four-point function is boosted relative to the square of the three-point function in the squeezed limit.
We derive all our results in two ways, using the barrier crossing formalism and the peak-background split method. 
In a companion paper~\cite{equivalence}, we prove that these two approaches are mathematically equivalent.}  
 \vspace{0.3cm}
 \hrule

\vspace{0.6cm}
\end{titlepage}

\tableofcontents

\newpage

\section{Introduction}

A central goal of modern cosmology is to uncover the physics that generated the primordial density perturbations and thereby seeded the large-scale structures (LSS) we see around us. 
The coherent nature of the cosmic microwave background (CMB) anisotropies suggests that the fluctuations were created at very early times, possibly during a period of inflation~\cite{inflation}.

One of the few observational probes that allows us access to the physics of that epoch is primordial non-Gaussianity~\cite{Komatsu:2009kd}.
At present, the best constraints on non-Gaussianity are coming from the CMB~(e.g.~\cite{Komatsu:2010hc}), but LSS is emerging as a promising complementary observable (e.g.~\cite{Desjacques:2010nn, Liguori:2010hx}).
Historically, the usefulness of LSS as a tool for early universe cosmology has been viewed with some suspicion, since non-linear evolution can itself produce significant non-Gaussianity even if the initial conditions were perfectly Gaussian.  
Disentangling any primordial non-Gaussianity from these late time effects always seemed like a messy business.
This attitude has changed somewhat when it was discovered that non-Gaussian initial conditions lead to a {\it scale-dependent} clustering of galaxies on large scales~\cite{Dalal:2007cu, Matarrese:2008nc}. 
In particular, it was shown that non-linear mode coupling induces a modulation of the local short-scale power $\sigma_8({\x})$ by the long-wavelength gravitational potential $\Phi({\x})$. 
This results in a biasing of halos (or galaxies) that is proportional to $\Phi$ rather than the dark matter density~$\delta$ (or $\nabla^2 \Phi$).
Crucially, the appearance of $\Phi$ rather than $\delta$ in the halo bias implies a specific form of scale-dependence that cannot be created dynamically (i.e.~by late time processes). This is the main reason that halo bias is such a robust probe of the initial conditions.

\begin{figure}[h!]
   \centering
       \includegraphics[width=9.cm]{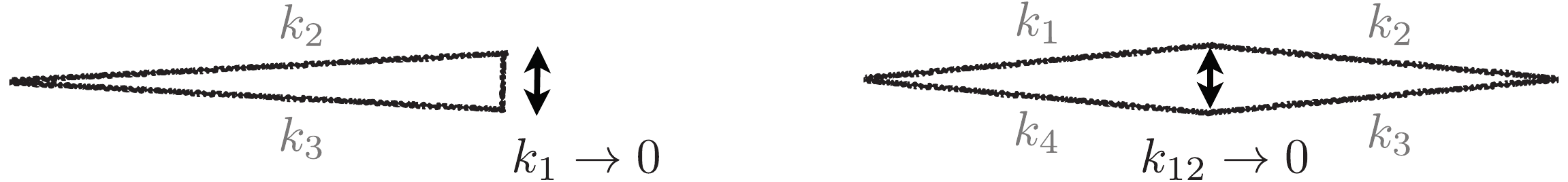}
   \caption{The squeezed limit of the three-point function, $k_1 \to 0$, gives the dominant contribution to the scale-dependent halo bias. A stochastic form of scale-dependent halo bias arises if the four-point function is large in the collapsed limit, $k_{12} \equiv |\k_1+\k_2| \to 0$.}
  \label{fig:34}
\end{figure}

In this paper, we study {\it stochastic} halo bias on large scales.
%Intuitively, we say that halo bias is stochastic if 
The term `stochastic' here refers to the fact that
the halo over-density is not 100\% correlated to the matter over-density on large scales,
i.e.~the halo-halo power spectrum $P_{\h \h}(k)$ is boosted relative to the matter-halo power spectrum $P_{\m\h}(k)$. Formally, %the bias is stochastic if 
this means that
\beq
P_{\h\h}(k) > b^2(k) P_{\m\m}(k) + \frac{1}{n_{\h}}\ ,
\eeq 
where $b(k) \equiv P_{\m\h}(k)/P_{\m\m}(k)$ is the halo bias, and $n_{\h}$ is the halo number density.
Large-scale stochastic bias arises in non-Gaussian models
when the small-scale power $\sigma_8({\x})$ varies from point to point, but in a way that isn't completely correlated with the local value of $\Phi({\x})$ and its derivatives.
This is most easily demonstrated in models with multiple fields, where the small-scale power may depend on fields that do not contribute to the (linearized) gravitational potential. 
%\db{Actually, in the previous sentence we mean only the Gaussian part of the gravitational potential. Shall we clarify?}%curvature perturbation. 
Our goal in this paper is to provide an understanding of the origin of stochastic bias in a model-independent way.  In the absence of significant isocurvature perturbations, all the relevant information must be encoded in the correlation functions of gravitational potential $\Phi$.  It will be useful to define
\begin{align}
\hat{f}_{\mathsmaller{\rm NL}}&\equiv \frac{1}{4} \lim_{k_1 \to 0} \frac{\xi^{(3)}_{\Phi}(\k_1,\k_2,\k_3)}{P_1 P_2} \ , \\
\hat{\tau}_{\mathsmaller{\rm NL}}  &\equiv \frac{9}{100} \lim_{k_{12}\to 0} \frac{\xi^{(4)}_\Phi(\k_1,\k_2,\k_3,\k_4)}{P_1 P_3 P_{12}} \ ,
\end{align}
where $\langle \Phi_{{\k}_1} \cdots \Phi_{{\k}_N} \rangle_{\rm c} \equiv (2\pi)^3 \xi^{(N)}_{\Phi}({\k}_1, \cdots \hskip -1pt , {\k}_n)\, \delta_{\rm D}({\k}_1 + \cdots + {\k}_n)$ and $P_i \equiv \xi_\Phi^{(2)}(k_i)$.
This parametrizes the amplitude of the three-point function in the {\it squeezed limit}, $k_1 \ll {\rm min}\{k_2,k_3\}$, and the amplitude of the four-point function in the  {\it collapsed limit}, $k_{12} \equiv |{\k}_1 + {\k}_2| \ll {\rm min}\{k_i\}$.
As we will show, stochastic bias arises if the `collapsed four-point function' is {\it not} equal to the square of the `squeezed three-point function', i.e.~if $\hat{\tau}_{\mathsmaller{\rm NL}}  \ne (\frac{6}{5}\hat{f}_{\mathsmaller{\rm NL}})^2$. 
There exists a well-known theoretical constraint on the relative size of $\hat{\tau}_{\mathsmaller{\rm NL}}$ and $(\frac{6}{5}\hat{f}_{\mathsmaller{\rm NL}})^2$.
If only a single field (which may or may not be the inflaton) generates the primordial curvature perturbation and its non-Gaussianity, then $\hat{\tau}_{\mathsmaller{\rm NL}} = (\frac{6}{5}\hat{f}_{\mathsmaller{\rm NL}})^2$ \cite{Byrnes:2006vq} and the biasing is non-stochastic. On the other hand, if multiple coupled fields generate the non-Gaussianity, then $\hat{\tau}_{\mathsmaller{\rm NL}}$ can be larger\footnote{No matter how the fluctuations were created, the parameters have to satisfy the Suyama-Yamaguchi inequality $\hat{\tau}_{\mathsmaller{\rm NL}}  \geq (\tfrac{6}{5}\hat{f}_{\mathsmaller{\rm NL}})^2$ \cite{SY} (see also \cite{Sugiyama:2011jt, Lewis:2011au, Smith:2011if, Quasi4,Kehagias:2012pd}). This is easy to understand: we can think of $\hat{f}_{\mathsmaller{\rm NL}}$ as a measure of the large-scale correlation between the potential $\Phi$ and
the locally measured small-scale power, $\hat{f}_{\mathsmaller{\rm NL}} \sim \langle \Phi_\ell \Phi_s^2\rangle/\langle \Phi_\ell^2 \rangle \langle \Phi_s^2\rangle$. On the other hand,   $\hat{\tau}_{\mathsmaller{\rm NL}}$  is a measure of the
large-scale variance in the small-scale power, $\hat{\tau}_{\mathsmaller{\rm NL}} \sim \langle \Phi_s^2 \Phi_s^2 \rangle_{\rm c}/\langle \Phi_\ell^2 \rangle \langle \Phi_s^2\rangle^2$. The inequality  $\hat{\tau}_{\mathsmaller{\rm NL}}  \geq (\tfrac{6}{5}\hat{f}_{\mathsmaller{\rm NL}})^2$ then
arises simply as the condition that the correlation coefficient between the small-scale power and $\Phi$ must be between $-1$ and $1$.} than $(\frac{6}{5}\hat{f}_{\mathsmaller{\rm NL}})^2$ \cite{Suyama:2010uj, ByrnesSource} and the biasing will be stochastic.
We will discuss classes of inflationary theories that predict precisely this kind of observational signature~\cite{Tseliakhovich:2010kf,Chen:2009zp, Baumann:2011nk}.
This provides the opportunity of using scale-dependent stochastic bias\footnote{We should note that in this paper we are interested in large-scale stochastic bias. On small scales, non-linear evolution and astrophysical processes can create local stochasticity, which is not relevant in our study.}  as a probe of any early universe physics associated with a boosted collapsed four-point function---just like the non-stochastic scale-dependent bias is a powerful probe of the squeezed three-point function.

More generally, we find that the large-scale non-stochastic bias %$b(k) = P_{\m\h}(k)/P_{\m\m}(k)$ 
can be written as a sum over $N$-point functions $\xi_\Phi^{(N)}(\k_1, \cdots \hskip -1pt,\k_N)$ evaluated in the squeezed limit $k_1 \ll \min\{k_2,\cdots \hskip -1pt,k_N\}$.\footnote{More precisely, $k_1$ is fixed to the large scale $k$ where we are computing the bias, and $k_2,\cdots \hskip -1pt ,k_N$ are integrated
over a broad range of scales near the halo collapse scale $k_\h \sim \rho_\m^{1/3} M^{-1/3}$.}
The stochastic bias, on the other hand, %$P_{\h\h}(k) - b^2(k) P_{\m\m}(k) - 1/n_\h$ can be written as
involves a double sum over $(M+N)$-point functions
$\xi_\Phi^{(M+N)}(\k_1,\cdots \hskip -1pt,\k_{M+N})$ evaluated in the collapsed limit $|\k_1 + \cdots + \k_M| \ll \min\{k_i\}$.
Stochastic bias arises if any collapsed $(M+N)$-point function is boosted relative to the product of the corresponding
squeezed $(M+1)$-point and $(N+1)$-point functions.
%However, 
In all physically interesting cases that we are aware of, this effect  %know of where stochastic bias arises, it 
is due to the collapsed four-point function
being boosted relative to the square of the three-point function (i.e.~the case $M=N=2$). Therefore, %and so 
we will generally interpret stochastic
bias as a probe of the collapsed four-point function. % throughout this paper.
The main result of this paper is a general pair of formulas, eqs.~(\ref{equ:PmhFinal}) and~(\ref{equ:PhhFinal}), 
for the non-stochastic and stochastic parts of the bias, for completely general non-Gaussian initial conditions parametrized by the $N$-point cumulants $\xi_\Phi^{(N)}(\k_1, \cdots\hskip -1pt, \k_N)$.

\vskip 6pt
The outline of the paper is as follows:
We will begin, in Section~\ref{sec:stochastic}, with a qualitative explanation of scale-dependent stochastic bias.
In Section~\ref{sec:barrier}, we will show how our intuitive understanding is borne out in the barrier crossing model of structure formation.
In Section~\ref{sec:examples}, we will illustrate these results with explicit examples.
In each case, we also derive our predictions in the peak-background split formalism. 
In a companion paper~\cite{equivalence}, we prove the mathematical equivalence of barrier crossing and peak-background split.
We present our conclusions in Section~\ref{sec:conclusions}.
Finally, Appendix~\ref{sec:convergence} discusses the convergence of the Edgeworth expansion for local non-Gaussianity.

\section{Stochastic Bias} \label{sec:stochastic}

Galaxies reside in dark matter halos.
For Gaussian initial conditions and at long wavelengths, the fluctuations in the density of halos $\delta_\h$ can be expressed as an expansion in the linear matter density field $\delta$. At linear order, the two are simply related by a numerical factor---the bias  $b_g$---i.e.~$\delta_\h = b_g \delta$.  
This simple bias relation gets modified for non-Gaussian initial conditions, due to a coupling between short and long-wavelength modes. The short modes determine the collapse of dark matter halos, while long modes modulate the density on large scales, effectively raising or lowering the threshold for the formation of collapsed objects. A non-zero three-point function  affects the variance of the short modes, leading to a dependence of the number density of halos on the amplitude of the long modes.
For local non-Gaussianity\footnote{In real space, local non-Gaussianity is parametrized as $\Phi({\x}) = \phi({\x}) + \fnl (\phi^2({\x}) - \langle \phi^2 \rangle)$, where $\phi$ is Gaussian.} this leads to a dependence of the halo density on the long-wavelength gravitational potential $\Phi$ rather than the matter density $\delta \propto\nabla^2 \Phi$. This leads to a characteristic {\it scale-dependence} in the bias relation, $\Delta b \propto k^{-2}$~\cite{Dalal:2007cu}.
It is this scale-dependence that allows us to trust the large-scale bias as a probe of initial conditions. Crucially, the dependence of the halo density on $\Phi$ is {not} something that could be mimicked by local dynamics. Dynamical processes don't care about the local value of the potential, but are only sensitive to tidal forces which are proportional to $\nabla^2 \Phi$ and $\dot \Phi$ (essentially this is a consequence of the equivalence principle). Any dependence of the small-scale power on $\Phi$ itself can therefore only come from the initial conditions.
This is what makes scale-dependent bias such a promising probe of early universe physics, despite all the astrophysical uncertainties associated with galaxy formation.
 
{\it Stochastic} bias arises whenever the density of halos is not 100\% correlated with the potential $\Phi$ or its derivatives.
In order to develop some intuition, we now give a schematic derivation of the effect. In the next section, we will upgrade this to a more formal analysis in the barrier crossing approach.
 If we assume that the primordial perturbations are adiabatic, then the formation of halos can only depend on local physics of the fluctuations.  Nevertheless, long-wavelength variations of the number of halos may depend, not only on the local value of the linear density field, but on all of its local correlation functions.  Assuming only locality, we may therefore write the local halo number density as
\beq
n_{\h}({\x}) = \bar n_\h(\delta({\x})  ; [ \delta^n]({\x})) \ ,
\eeq
where $[ ... ]$ denotes an average over a small region of characteristic size $\ell$ that is centered around~$\x$.  Long-wavelength fluctuations in the number of halos can then be understood as a Taylor expansion,
\beq
\delta_{\h}({\x}) \equiv \frac{\delta n_\h}{\bar n_\h} = b_g \delta({\x}) +  \beta [ \delta^2 ]({\x}) + \cdots \ , \label{equ:Taylor}
\eeq
where $b_g$ is the Gaussian bias and 
\beq
\beta \equiv   \frac{\partial\ln n_\h}{\partial   {[ \delta^2 ]}} \ .
\eeq  
It is easy to see (e.g.~by splitting all fields into long and short modes), that for local non-Gaussianity the short-scale power is modulated by the gravitational potential, $ [ \delta^2 ]\approx [ \delta^2 ]_g \,  (1 + 4 \fnl \Phi({\x}))$. 
%\db{Shall we clarify $\overline{ \langle \delta^2 \rangle}$?}
%\kms{Sounds good to me; also suggest writing $[ \delta^2 ] \approx \overline{ \langle \delta^2 \rangle} (1 + 4 \fnl \Phi({\x}))$}
This is the origin of scale-dependent bias in local non-Gaussianity.

Using the expansion (\ref{equ:Taylor}), we can also evaluate correlation functions between two spatially separated points ${\x}$ and ${\x}'$.  We use a prime to indicate that fields are evaluated at ${\x}'$, while fields without a prime are evaluated at ${\x}$.
The matter-halo correlation, in a large region of size $L \gg \ell$, then is
\beq\label{eqn:simplehm}
\frac{\langle \delta_\h \delta' \rangle}{\langle \delta \delta' \rangle} = b_g + \beta \frac{\langle [ \delta^2 ] \hskip 1pt \delta' \rangle}{\langle \delta \delta' \rangle} + \cdots \ ,
\eeq
while the halo-halo correlation is
\beq\label{eqn:simplehh}
\frac{\langle \delta_\h \delta_\h{}'  \rangle}{\langle \delta \delta' \rangle} = b_g^2 + 2 b_g \beta \frac{\langle [ \delta^2 ] \hskip 1pt \delta' \rangle}{\langle \delta \delta' \rangle} + \beta^2 \frac{\langle [ \delta^2 ] [ \delta^2 ]' \rangle}{\langle \delta \delta' \rangle} + \cdots \ .
\eeq
This leads to the possibility that the bias inferred from $\langle \delta_\h \delta \rangle$ is not equal to the bias inferred from $\langle \delta_\h \delta_\h{}' \rangle $.
We characterize this so-called stochasticity of the halo bias by the following parameter\footnote{In practice, we also have to subtract shot noise contributions from $\langle \delta_\h \delta_\h{}' \rangle$ and $\langle \delta_\h \delta' \rangle$---see \S\ref{ssec:barrier_crossing}.}
\beq
r \equiv \frac{\langle \delta_\h \delta_\h{}' \rangle}{\langle \delta \delta' \rangle} - \left( \frac{\langle \delta_\h \delta' \rangle}{\langle \delta \delta' \rangle}\right)^2  \ . \label{equ:rdef}
\eeq
Using eqs.~(\ref{eqn:simplehm}) and (\ref{eqn:simplehh}), we find
\beq
r = \beta^2 \left[ \frac{\langle [ \delta^2 ] [ \delta^2 ]' \rangle}{\langle \delta \delta' \rangle}  - \left( \frac{\langle [ \delta^2 ] \hskip 1pt \delta' \rangle}{\langle \delta \delta' \rangle}\right)^2 \, \right] + \cdots\ . \label{equ:rr}
\eeq
This simple argument gives reliable intuition for the origin of stochasticity.  Specifically, we see that if a local variation in the amplitude of $[ \delta^2 ]({\x})$ is uncorrelated with $\delta({\x}')$, then there is no extra contribution to the bias in eq.~(\ref{eqn:simplehm}).  Nevertheless, the halo-halo correlation function in eq.~(\ref{eqn:simplehh}) can still be modified by long-wavelength variations in $[ \delta^2 ]({\x})$.  
Moreover, the result (\ref{equ:rr}) makes it clear that stochasticity arises from a non-trivial four-point function of the primordial potential.  In fact, the real space correlation function $\langle [ \delta^2] ({\x}) [ \delta^2  ] ({\x}') \rangle$ relates to the collapsed limit of the four-point function in Fourier space, i.e.~$\lim_{|{\k}_{1}+{\k}_2| \to 0} \langle \Phi_{{\k}_1}  \Phi_{{\k}_2}   \Phi_{{\k}_3}  \Phi_{{\k}_4}\rangle$.

\section{ Predictions from Barrier Crossing}
\label{sec:barrier}

In this section, we give a formal derivation of stochastic bias using the classic barrier crossing method of Press and Schechter \cite{Press:1973iz}. 
Our goal is to obtain an expression for the stochasticity coefficient~(\ref{equ:rdef}) in terms of the cumulants of the smoothed density field. These in turn can be related to $N$-point functions of the primordial potential and hence contain information about the initial conditions.

\subsection{Definitions and Notation}

We begin with some basic definitions and a description of our notation.
Let $\hat \delta(\x, z)$ denote the linear density field (to be distinguished by the hat from the non-linear density field $\delta$). 
The linearized Poisson equation relates $\hat \delta$ to the primordial potential $\Phi$,
\beq
\hat \delta_{\k}(z) = \alpha(k,z) \Phi_{\k}\ ,
\eeq
where
\beq \alpha(k,z) \equiv \frac{2}{3} \frac{k^2}{\Omega_m H_0^2} \, T(k) D(z) \ .\eeq
Here, $T(k)$ is the matter transfer function normalized such that $T(k) \rightarrow 1$ as $k \rightarrow 0$ and $D(z)$ is the linear growth factor (as function of redshift $z$), normalized so that $D(z) = (1+z)^{-1}$ in matter domination.  For notational simplicity, we will from now on suppress the redshift argument from all quantities.
We use $\delta_\M(\x)$ for the linear field smoothed with a top-hat window function with radius\footnote{The smoothing scale $R_\M$ corresponds to the comoving size of halos of mass $M$ in Lagrangian space.} $R_\M \equiv (3M/4 \pi \bar{\rho}_\m)^{1/3}$, so that  
%\db{I added $z$'s to the arguments of various functions here:} \db{Do we need to comment that we soon drop the $z$ arguments, e.g. in $\delta_\M$ and $\alpha_\M$?}
\beq
\delta_\M({\x}) = \int_{\k} e^{- i{\k} \cdot {\x}} \, W_\M(k) \hat \delta_{\k}  =  \int_{\k} e^{- i{\k} \cdot {\x}} \, \alpha_\M(k) \Phi_{\k} \ ,
\eeq
where $\int_\k\, (\cdot) \equiv \int \frac{\d^3 \k}{(2\pi)^3}\, (\cdot)$,
\beq\label{eqn:tophat}
W_\M(k) \equiv 3\, \frac{\sin(kR_\M) - kR_\M \cos(kR_\M)}{(kR_\M)^3}\ ,
\eeq
and $\alpha_\M(k) \equiv W_\M(k) \alpha(k)$. 
Let $\sigma_\M = \langle \delta^2_\M \rangle^{1/2}$  be the rms amplitude of the smoothed density field, and $\kappa_n(M)$ be its $n$-th non-Gaussian cumulant, 
\beq
\kappa_n(M) = \frac{\langle \delta_{\M}^n \rangle_{\rm c}}{\sigma_{\M}^n} \ ,  \label{eq:kappa_def}
\eeq
where the subscript `${\rm c}$' indicates the use of a connected correlation function.
Since $\delta_\M$ and $\sigma_\M$ are defined via linear theory, $\kappa_n(M)$ is independent of redshift. 
Similar definitions apply to the unsmoothed field $\hat \delta$, in which case we denote the variance and cumulants by $\hat \sigma$ and $\kappa_{\hat n}$.

Ultimately, we will be interested in two-point clustering statistics.
Let ${\x}$ and ${\x}'$ be two points separated by a distance $r \equiv |{\x}-{\x}'|$. Moreover, let a prime indicate that the field is evaluated at~${\x}'$, e.g.~$\delta_{\M}' \equiv \delta_{\M}({\x}')$. Fields without a prime are evaluated at ${\x}$.
The joint cumulants are then defined by  
\begin{align}
\kappa_{\hat m,n}(r,M) &\ \equiv\ \frac{\langle \hat \delta^m (\delta_{\M}')^n\rangle_{\rm c}}{\hat \sigma^m\sigma_{\M}^n}\ , \label{equ:kk1} \\
\kappa_{m,n}(r,M,\bar M) &\ \equiv\ \frac{\langle (\delta_{\M})^m (\delta_{\bar \M}')^n\rangle_{\rm c}}{\sigma_{\M}^{m} \sigma_{\bar \M}^n}  \ . \label{equ:kk2}
\end{align}
These cumulants can be related to $N$-point functions of the gravitational potential,
\beq
\langle \Phi_{{\k}_1}  \Phi_{{\k}_2} \cdots  \Phi_{{\k}_N}\rangle_{\rm c} = (2\pi)^3 \delta_{\rm D}({\k}_{1 2\dots N}) \, \xi_\Phi^{(N)}({\k}_1, {\k}_2, \dots, {\k}_N)\ ,
\eeq
where ${\k}_{1 2\dots N} \equiv {\k}_{1} + {\k}_{2} + \cdots + {\k}_{N}$.

\subsection{Edgeworth Expansions}
\label{sec:edge}

The probability density functions (PDFs) of weakly non-Gaussian random variables have well-defined Edgeworth expansions (for a review see e.g.~\cite{Bernardeau:2001qr}). 
Consider first the variables $\delta_{\M}$ and $\delta'_{\M}$.
It will be convenient to define the rescaled fields 
\beq
\nu \equiv  \frac{\delta_{\M}}{\sigma_{\M}} \qquad {\rm and} \qquad \nu \hskip 1pt{}' \equiv \frac{\delta_{\M}'}{\sigma_\M}\ , 
\eeq
with 
$\langle \nu \rangle = \langle \nu \hskip 1pt{}' \rangle = 0$ and $\langle \nu^2 \rangle = \langle (\nu \hskip 1pt{}')^2 \rangle = 1$. 
The cumulants in eqs.~(\ref{eq:kappa_def}) and (\ref{equ:kk2}) then become $\kappa_n = \langle \nu^n\rangle_{\rm c}$ and $\kappa_{m,n}=\langle \nu^m (\nu \hskip 1pt{}')^n\rangle_{\rm c}$. 
The Edgeworth expansion for the marginal PDF is
\begin{align}
p(\nu) &= \exp \left( \sum_{n \ge 3} \frac{(-1)^n}{n!}  \kappa_n  \frac{\partial^n}{\partial \nu^n} \right) p_g(\nu) \ , \qquad {\rm where} \quad p_g(\nu) \equiv \frac{1}{\sqrt{2\pi}} e^{- \frac{1}{2} \nu^2}\ .
\end{align}
The first few terms can be written as
\begin{align}
p(\nu)  & \ = \ \left( 1+ \frac{\kappa_3}{3!} H_3(\nu) + \frac{\kappa_4}{4!} H_4(\nu) + \cdots \right) p_g(\nu) \ , \label{equ:p1}
\end{align}
where the functions $H_n(\nu)$ are Hermite polynomials
\beq
H_n(\nu) \equiv (-1)^n e^{\frac{1}{2} \nu^2} \frac{d^n}{d\nu^n} e^{-\frac{1}{2} \nu^2 } \ .
\eeq
Similarly, the Edgeworth expansion for the joint PDF is%\footnote{In Appendix~\ref{sec:convergence}, we discuss the convergence properties of this Edgeworth expansion.}
\begin{align}
p(\nu,\nu \hskip 1pt{}') = \exp \left( \kappa_{1,1} \frac{\partial^2 }{\partial \nu \partial \nu \hskip 1pt{}'} + \sum_{m+n \ge 3} \frac{(-1)^{m+n}}{m! n!} \kappa_{m,n} \frac{\partial^{m+n}}{\partial \nu^m \partial (\nu \hskip 1pt{}')^n} \right) p_g(\nu) p_g(\nu \hskip 1pt{}')\ . \label{equ:edge1}
\end{align}
In Appendix~\ref{sec:convergence}, we discuss the convergence properties of this expansion.
In the next section, we will use it to compute halo-halo correlations. 

The matter-halo case is completely analogous: to construct the joint PDF of the variables $\hat\delta$ and $\delta'_{\M}$, we define
rescaled variables $\hat\nu = \hat\delta/\hat\sigma$ and $\nu' = \delta'_{\M}/\sigma_{\M}$.  The joint PDF $p(\hat\nu,\nu')$ is then given
by the Edgeworth series~(\ref{equ:edge1}) with the cumulant $\kappa_{m,n}$ replaced by $\kappa_{\hat m,n}$.

\subsection{Barrier Crossing}
\label{ssec:barrier_crossing}

In the simplest version of the barrier crossing formalism~\cite{Press:1973iz}, halos of mass $\geq M$ are identified with regions where the \textit{linearly} evolved smoothed density field exceeds a constant threshold value~$\delta_c$ for collapse. The halo number density $n_\h(\x)$ is then given by
\beq
n_\h({\x}) = 2\hskip 1pt \frac{\bar{\rho}_{\rm m}}{M} \, \Theta(\delta_\M({\x}) - \delta_c)\ ,   \label{eq:nh_barrier}
\eeq
with $\Theta$ the Heaviside step function.
It has been shown numerically that $\delta_c \approx 1.42$ produces good results~\cite{Grossi:2009an}, but for our analytical calculations we don't need to specify a particular value for $\delta_c$.
The fraction of space occupied by regions above the collapse threshold is 
\beq
f(M)  
=
 \int_{\nu_c}^{\infty} [\d \nu]\ p(\nu)\ ,
\eeq
where %$\nu \equiv \delta_\M/ \sigma_\M$ and 
$\nu_c(M) \equiv \delta_c/\sigma_\M$.
Using the Edgeworth expansion (\ref{equ:p1}), we find\footnote{In our notation the halo mass function is $d n_\h/d M = -\hskip 1pt \bar{\rho}_\m/M (d f / d M)$.}
\beq
f(M) = \frac{1}{2} \, {\rm erfc} \left( \frac{\nu_c}{\sqrt{2}} \right) + p_g(\nu_c) \left[ \frac{\kappa_3(M)}{3!} H_2(\nu_c) + \frac{\kappa_4(M)}{4!} H_3(\nu_c) + \cdots \right]\ . \label{equ:x0}
\eeq
When interpreting calculations in the barrier crossing model, it must be kept in mind that eq.~(\ref{eq:nh_barrier}) for $n_\h(\x)$ is the number density of halos in {\em Lagrangian} space.  Our convention throughout this paper is that the power spectra $P_{\m\h}(k)$ and $P_{\h\h}(k)$ are always computed in Lagrangian space.  In particular, $b_g$ denotes the Lagrangian bias.  The relevant quantity to compare to observations or simulations is the {\it Eulerian} bias 
which, to lowest order, is given by $b_g^{\mathsmaller{\rm E}} = 1 + b_g$.
%, since the Lagrangian and Eulerian halo fields are related by $\delta_\h^{\mathsmaller{\rm E}} = \delta_\h^{\mathsmaller{\rm L}}) + \delta$. \db{We haven't defined $\delta_\ell$ yet.}

The barrier crossing model also neglects shot noise contributions which arise from the finite halo number
density $n_\h$.  Throughout this paper, $P_{\h\h}$ always denotes the halo-halo power spectrum after subtracting
the shot noise contribution $1/n_\h$.
(There are also shot noise, or one-halo, contributions to the matter-halo power spectrum $P_{\m\h}$, which are usually negligible, but are a leading source of stochastic bias in the Gaussian case \cite{Hamaus:2010im,Smith:2010gx}.)

\subsubsection{Matter-Halo Correlations}

The correlation between the halo field at
$\x'$ and the dark matter field at $\x$ is given by
\begin{align}
\hat \xi(r,M) &\ =\ \frac{M}{2 \bar{\rho}_\m} \int_{- \infty}^\infty [\d \hat \delta] \int_{-\infty}^\infty [\d \delta_{\M}']  \, \hat \delta({\x}) \, n_\h(\x')\, p(\hat \delta, \delta_{\M}') \ . \label{equ:hatxi1}
\end{align}
In the rescaled variables $\hat \nu \equiv \hat \delta(\x)/\hat \sigma$ and $\nu \hskip 1pt{}' \equiv \delta_\M(\x')/\sigma_\M$, this becomes
\begin{align}
\hat \xi(r,M) &\ =\ \hat \sigma \int_{- \infty}^\infty [\d \hat \nu] \int_{\nu_c}^\infty [\d \nu \hskip 1pt{}']  \, \hat \nu \, p(\hat \nu,\nu \hskip 1pt{}') \ . \label{equ:hatxi2}
\end{align}
It will be convenient to work in momentum space via $\hat \xi(k, M) = \int \d^3 {\r} \, e^{i {\k}\cdot {\r} } \hat \xi(r, M)$.  To describe the correlations of halos in the mass bin $[M,M+\d M]$, we take derivatives with respect to $M$.
The matter-halo power spectrum is then given by %\footnote{Before this result can be compared to simulations and/or observations, shot noise has to be subtracted from the latter. \label{foot:5}}
\beq
P_{\rm mh}(k,M) = \frac{d \hat \xi(k,M)}{d M} \, \left(\frac{d f(M)}{d M}\right)^{-1}\ . \label{equ:Pmh}
\eeq
%where the second term is due to shot noise and $f_\h$ is the fraction of dark matter in halos.  
To compute the correlation function~(\ref{equ:hatxi2}), we
substitute the Edgeworth expansion (\ref{equ:edge1}) for $p(\hat \nu, \nu \hskip 1pt{}')$.
Only terms with exactly one $\hat \nu$-derivative survive the integration, and we therefore find
\beq
\hat \xi(k,M) =  \hat \sigma\, p_g(\nu_c) \left[ \kappa_{\hat 1,1} + \frac{H_1(\nu_c)}{2!} \kappa_{\hat 1,2} +  \frac{H_2(\nu_c)}{3!} \kappa_{\hat 1,3} + \frac{H_3(\nu_c)}{3!} \kappa_{\hat 1,1} \star \kappa_3 + \cdots \right] \ , \label{equ:hatxi3}
\eeq
where $\star$ denotes a convolution.  
We see that the matter-halo correlations, or equivalently the non-stochastic part of the halo bias, only depend on the following cumulants 
%\db{check}
\beq
\kappa_{\hat 1,n}(k,M) = \hat \sigma^{-1} \sigma_{\M}^{-n}  \left(\, \prod_{i=1}^n \int_{{\q}_i} \alpha_\M(q_i) \right) \alpha(k) \langle \Phi_{\k} \Phi_{{\q}_1} \cdots \Phi_{{\q}_n} \rangle_{\rm c} \ .
\eeq
Moreover, we note that the large-scale limit, $\lim_{k\to 0} \kappa_{\hat 1,n}(k,M) $, is determined by the {squeezed limit} of the primordial $(n+1)$-point function \cite{Desjacques:2011mq}, 
\beq
\lim_{k\to 0}  \xi_\Phi^{(n+1)}({\k}, {\q}_1, \cdots \hskip -1pt ,{\q}_n ) \ .
\eeq
The explicit form of the cumulants $\kappa_{\hat 1,n \ge 2}$ depends on the type of non-Gaussianity. We compute some examples in Section~\ref{sec:examples}.

Keeping only linear terms\footnote{In Appendix~\ref{sec:convergence}, we explain that the lowest order cumulants usually dominate and that products of cumulants are suppressed.} in eq.~(\ref{equ:hatxi3}), we get \beq 
\hat \xi(k,M) = p_g(\nu_c) ( \kappa_{\hat 1, 1} \hat \sigma \sigma_\M) \left[\frac{1}{\sigma_\M} + \sum_{n=2}^\infty \frac{H_{n-1}(\nu_c)}{n!} \, f_{\hat 1,n} + \cdots \right] \ ,  \label{eq:xi_mh_linear}
\eeq
where we defined 
\beq
f_{\hat 1, n}(k,M) \equiv \frac{\kappa_{\hat 1, n}(k,M)}{\kappa_{\hat 1, 1}(k,M) \, \sigma_\M}\ . \label{equ:f1n}
\eeq
It was convenient to factor out the Gaussian term $\kappa_{\hat 1,1}  \hat \sigma \sigma_\M$, since at long wavelengths it becomes the matter power spectrum
\beq
\kappa_{\hat 1, 1} \, \hat \sigma \sigma_\M =  \int_{\q} \alpha(k) \alpha_\M(q) \langle \Phi_{\k} \Phi_{\q} \rangle =  W_\M(k)  P_{\m \m}(k) \ \xrightarrow{k\to 0} \ P_{\m \m}(k)\ .
\eeq
% Here and throughout the paper, we are implicitly dropping the one-halo contribution to $P_{\m\m}$ and identify $P_{\m\m}$ with the linear matter power spectrum $\hat P_{\m \m}$.  Although the one-halo term does contribute significantly to the stochastic bias for Gaussian initial conditions \cite{Smith:2010gx}, it will be negligible in the models of interest.   
Evaluating eq.~(\ref{equ:Pmh}), we find
\begin{align}
P_{\rm mh}(k)  &\ \xrightarrow{k\to 0} \  P_{\m \m}(k) \left[ \, b_g + \sum_{n=2}^\infty \left( \beta_n +  \tilde \beta_n \frac{d }{d \ln \sigma_\M} \right) f_{\hat 1,n} + \cdots  \right]\ , \label{equ:PmhFinal}
\end{align}
where %\footnote{Note that the thresholding was performed on the {\it Lagrangian} density field. The relevant quantity to compare to observations or simulations is the {\it Eulerian} bias which, to lowest order, is given by $b_g^{\rm E} = 1 + b_g$, since the Lagrangian and Eulerian halo fields are related by $(1+\delta_\h^{\rm E}) = (1+\delta_\ell)(1+\delta_\h^{\rm L})$.}
\begin{align}
b_g &\equiv  \frac{1}{\sigma_\M}\frac{\nu_c^2 -1}{\nu_c}\quad , \quad \beta_{n} \equiv \frac{H_n(\nu_c)}{n!} \quad {\rm and} \quad \tilde \beta_{n} \equiv \frac{H_{n-1}(\nu_c)}{n!\,\nu_c}\ .
\end{align}
The ellipses in eq.~(\ref{equ:PmhFinal}) stand for terms that are non-linear in the cumulants.
For local non-Gaussianity, the derivative terms $df_{\hat 1,n}/d\ln \sigma_\M $ will be negligible, but in principle, we can keep them (and sometimes we have to). 
%Up to fourth order (and not showing the derivative terms), we have
%\begin{align}
%P_{\rm mh}(k) &\ \xrightarrow{k\to 0} \  P_{\m \m}(k) \left[ \, b_g + \beta_2\, f_{\hat 1,2}  + \beta_3\, f_{\hat 1,3} + \cdots  \right]\ . \label{equ:PmhFinal2}
%\end{align}
%\kms{Get rid of the preceding equation?  It doesn't seem to add much.  If we do show it, I think we should include derivative terms.}

%\db{I modified our comments on Desjacques et al. a bit. Does everybody agree?}
Our expression~(\ref{equ:PmhFinal}) agrees with the general formula for the non-stochastic bias given in \cite{Desjacques:2011mq}; however,
ref.~\cite{Desjacques:2011mq} implicitly found that non-Gaussianity cannot generate large-scale stochastic bias. In the next section, we will find the
opposite conclusion. The disagreement is easy to understand: Ref.~\cite{Desjacques:2011mq} claims after their eq.~(40) that contributions to $P_{\h\h}$
from cumulants $\kappa_{m,n}$ with $m,n\ge 2$ must approach a constant as $k\rightarrow 0$. This is not true for general non-Gaussian initial conditions and exceptions to that statement are precisely what causes the effects discuss in this paper.

\subsubsection{Halo-Halo Correlations}
\label{sec:hh}

Next, we consider the correlation between the halo fields at $\x$ and $\x'$,
\begin{align}
\xi(r,M,\bar M) &\ =\  \frac{M \bar M}{4 \bar{\rho}_\m^2} \int_{-\infty}^\infty [\d \delta_{{\M}}] \int_{-\infty}^\infty [\d \delta_{{\bar\M}}']  \, n_\h(\x) n_\h(\x')\, p(\delta_{{\M}}, \delta_{{\bar \M}}') \ .
\end{align}
In the rescaled variables $\nu \equiv \delta_\M(\x)/\sigma_\M$ and $\nu \hskip 1pt{}' \equiv \delta_{\bar \M}(\x')/\sigma_{\bar \M}$,  this becomes
\beq
\xi(r, M, \bar M) \ = \ \int_{\nu_c}^\infty [\d \nu] \int_{\bar \nu_c}^\infty [\d \nu \hskip 1pt{}'] \ p(\nu,\nu \hskip 1pt{}')\ .  \label{equ:xi1}
\eeq
Notice that, in principle, we have allowed for two distinct mass thresholds, $M$ and $\bar M$.
The power spectrum of halos in the mass bins $[M, M+\d M]$ and $[\bar M, \bar M + \d \bar M]$ then is
\beq
P_{\h \bar \h}(k) = \frac{d^2 \xi(k,M,\bar M)}{d M d \bar M} \, \left(\frac{d f(M)}{d M} \frac{d f(\bar M)}{d \bar M}\right)^{-1} \ .
\eeq
For simplicity, we will restrict the following presentation to correlations of equal mass halos, $M= \bar M$. The power spectrum for a narrow mass bin around $M$ is then given by %\footnote{See footnote~\ref{foot:5}.}
\beq
P_{\h \h}(k) = \frac{d^2 \xi(k,M, \bar{M})}{d M d \bar{M}} \bigg |_{\bar{M} = M} \, \left(\frac{d f(M)}{d M}\right)^{-2}  \ .\label{equ:Phh}
\eeq
To compute the correlation function~(\ref{equ:xi1}), we
substitute the Edgeworth expansion (\ref{equ:edge1})
 for $p(\nu,  \nu \hskip 1pt {}')$,
\begin{align}
\xi(k,M, \bar M) &= p_g(\nu_c) p_g(\bar \nu_c) \left[ \kappa_{1, 1} +  \frac{1}{2} \big( \kappa_{2, 1} H_1(\nu_c) + \kappa_{ 1, 2} H_1(\bar \nu_c) \big) +  \frac{1}{6} \big( \kappa_{3, 1} H_2(\nu_c) + \kappa_{ 1, 3} H_2(\bar \nu_c) \big) \nonumber \right. \\
 &\left. \hspace{4cm}+  \frac{1}{4} \kappa_{2,2} H_1(\nu_c) H_1(\bar \nu_c) +  \frac{1}{2} \kappa_{1,1} \star \kappa_{1,1} + \ \cdots \right] \ . \label{equ:xi2}
\end{align}
The form of higher-order cumulants, such as $\kappa_{1,2}$, $\kappa_{1,3}$ and $\kappa_{2,2}$, again depends on the type of non-Gaussianity. 
We compute some examples in Section~\ref{sec:examples}.
% Of particular importance is the cumulant
% \beq
% \kappa_{2,2}(k,M,\bar M) \ \xrightarrow{k\to 0} \  \int \limits_{{\q}_1} \int \limits_{{\q}_2} \frac{\alpha_{\M}^2(q_1)}{\sigma_{\M}^2} \frac{\alpha_{\bar \M}^2(q_2)}{\sigma_{\bar \M}^2} \, \xi^{(4)}_\Phi({\q}_1, {\k} - {\q}_1, {\q}_2, {\q}_2 - {\k})\ ,
% \eeq
% which depends on the collapsed limit of the four-point function,
% \beq
% \lim_{k_{12} \to 0} \xi^{(4)}_\Phi({\k}_1, {\k}_2, {\k}_3, {\k}_4)\ .
% \eeq

Keeping only the terms linear in $\kappa_{m,n}$ (this approximation will be justified in Appendix~\ref{sec:convergence}) in eq.~(\ref{equ:xi2}), we find 
\begin{align}
\xi(k,M,\bar M) &\ =\ p_g(\nu_c) p_g(\bar \nu_c) (\kappa_{1,1} \sigma_\M \sigma_{\bar \M})\left[ \frac{1}{\sigma_\M \sigma_{\bar \M}} +  \sum_{n=2}^\infty \left(\frac{1}{\sigma_\M}  \frac{H_{n-1}(\bar \nu_c) }{n!}  \, f_{ 1, n}  + \frac{1}{\sigma_{\bar \M}}   \frac{H_{n-1}(\nu_c) }{n!}  \, f_{ n, 1}\right)  \nonumber \right. \\
& \left. \hspace{5.1cm} +\sum_{m=2}^\infty \sum_{n=2}^\infty  \frac{H_{m-1}(\nu_c)}{m!}  \frac{H_{n-1}(\bar \nu_c)}{n!} f_{m,n}\ +\ \cdots \ \right] \ , \label{equ:xi3}
\end{align}
where %\db{Should it be $f_{1,n}(k,M,\bar M)$?}
\begin{align}
f_{1,n}(k,M,\bar M) &\equiv \frac{\kappa_{1,n}(k,M,\bar M)}{\kappa_{1,1}(k,M,\bar M) \sigma_{\bar\M}} \ \ \ \, \, \qquad \mbox{for $n \ge 1$}\ , \\
f_{m,n}(k,M,\bar M) & \equiv \frac{\kappa_{m,n}(k,M,\bar M)}{\kappa_{1,1}(k,M,\bar M) \sigma_{\M} \sigma_{\bar \M}} \qquad \mbox{for $m,n \ge 2$}\ .
\end{align}
We again factored out the Gaussian contribution, $\kappa_{1,1} \, \sigma_\M \sigma_{\bar \M}  \ \xrightarrow{k\to 0} \ P_{\m \m}(k)$.
Note that $f_{1,n} = f_{\hat 1,n}$ in the large scale limit $k \ll R_{\M}^{-1}$, where $f_{\hat 1,n}$ was defined in eq.~(\ref{equ:f1n}).
Substituting (\ref{equ:xi3}) into (\ref{equ:Phh}), we get
\begin{align}
P_{\h \h}(k) &\ \xrightarrow{k\to 0} \  P_{\m \m}(k) \left[\,  b_g^2 + 2 b_g \sum_{n=2}^\infty \left( \beta_n + \tilde \beta_n \frac{\partial}{\partial \ln \sigma_\M} \right)  f_{1,n}\right. \nonumber \\
& \hspace{2cm} \left. + \sum_{m=2}^\infty \sum_{n=2}^\infty 
   \left( \beta_m + \tilde\beta_n\, \frac{\partial}{\partial\ln\sigma_{\M}} \right)
   \left( \beta_m + \tilde\beta_n\, \frac{\partial}{\partial\ln\sigma_{\bar\M}} \right) f_{m,n} +\ \cdots\ \right]\ . \label{equ:PhhFinal}
\end{align}
Note that, while in the end we always take $M=\bar M$ in this paper, $M$ and $\bar M$ are independent variables
when calculating partial derivatives of $f_{m,n}(M,\bar M)$.

\subsection{Stochastic Halo Bias}

We now combine the above results to evaluate the stochasticity coefficient
\beq
r \equiv \frac{P_{\rm hh} }{P_{\rm mm}} - \left( \frac{P_{\rm mh}}{P_{\rm mm}}\right)^2 \ ,
\eeq
where, as usual, it is understood that shot noise is subtracted from $P_{\h \h}$  and $P_{\m \h}$.
Substituting eqs.~(\ref{equ:PmhFinal}) and (\ref{equ:PhhFinal}), we find %\db
\beq
r \xrightarrow{k\to 0} \sum_{m=2}^\infty \sum_{n=2}^\infty
   \left( \beta_m + \tilde\beta_m \frac{\partial}{\partial\ln\sigma_{\M}} \right)
   \left( \beta_n + \tilde\beta_n \frac{\partial}{\partial\ln\sigma_{\bar\M}} \right) f_{m,n}
  - \left[ \, \sum_{n=2}^\infty \left( \beta_n + \tilde\beta_n \frac{\partial}{\partial\ln\sigma_\M} \right) f_{n,1} \right]^2 \ . \label{equ:master}
\eeq
We note that cumulants $\kappa_{m,n}(k)$ with $m,n\ge 2$ contribute to the halo-halo power spectrum but not the matter-halo power spectrum~(\ref{equ:PmhFinal}),
so stochastic halo bias is sourced by these cumulants.  These cumulants can be written in terms of the $(m+n)$-point functions of the gravitational potential,
\begin{align}
\kappa_{m,n}(k,M,\bar M) & \ \xrightarrow{k\to 0}\ \frac{1}{\sigma_{\M}^m \sigma_{\bar\M}^n} \left( \prod_{i=1}^{m-1} \int_{\q_i} \alpha_\M(q_i) \right) \left( \prod_{j=1}^{n -1} \int_{\q_j'} \alpha_{\bar \M}(q_j') \right) \alpha_\M(q) \,\alpha_{\bar \M}(q')
\nonumber \\
& \hspace{1cm}\ \ \times 
  \xi_\Phi^{(m+n)}\left( \q_1,\cdots \hskip -1pt ,\q_{m-1},- \q+\k,\q'_1,\cdots \hskip -1pt,\q'_{n-1},-\q' -\k \right) \ ,
\end{align}
where $\q \equiv \sum_{i=1}^{m-1}\q_i$ and $\q' \equiv \sum_{j=1}^{n-1}\q_j'$.
We see that, in general, large-scale stochastic bias arises whenever an $(m+n)$-point function $\xi_\Phi^{(m+n)}(\k_1,\cdots \hskip -1pt,\k_{m+n})$ 
is boosted in the collapsed limit $|\sum_{i=1}^m \k_i| \rightarrow 0$, relative to the product of the corresponding squeezed $(m+1)$-point and $(n+1)$-point
functions. 
In the next section, we will compute eq.~(\ref{equ:master}) for a few interesting examples.
In most cases, we will get stochastic bias from the case $m=n=2$,
i.e.~a collapsed four-point function $\lim_{|\k_1+\k_2|\rightarrow 0} \xi^{(4)}_\Phi(\k_1,\k_2,\k_3,\k_4)$ 
which is larger than the square of the squeezed three-point function $\lim_{k_1\rightarrow 0} \xi^{(3)}_\Phi(\k_1,\k_2,\k_3)$.

\newpage
\section{Examples}
\label{sec:examples}

In this section, we discuss several physical mechanisms that lead to 
 stochastic halo bias.
For each example, we will derive the result in two different ways:
\begin{enumerate}
\item[1)] using a peak-background split (PBS) method;
\item[2)] using the barrier crossing analysis of the previous section.
\end{enumerate}
We demonstrate explicitly that both approaches lead to the same answers.

\subsection{$\tau_{\rm NL}$ Cosmology} 
\label{sec:tnl}

A simple phenomenological way to get a boosted collapsed limit for the four-point function is the following generalization of the local ansatz to multiple fields
\beq
\Phi = A_i \phi_i + B_{ij} \left( \phi_i \phi_j - \langle \phi_i \phi_j \rangle \right)\ ,
\eeq
with the Einstein summation convention understood.
This structure arises, for example, in the curvaton model of \cite{Tseliakhovich:2010kf} (see also \cite{Smith:2010gx}), 
\beq\label{eqn:taunlmodel}
\Phi = \phi + \psi + \fnl(1+\Pi)^2 \, (\psi^2 - \langle \psi^2 \rangle)\ , \qquad {\rm where} \quad \frac{P_\phi}{P_\psi} \equiv \Pi\ .
\eeq
Here, $\phi$ and $\psi$ are uncorrelated Gaussian random fields with power spectra that are proportional to each other. 
The three- and four-point functions take the local form
\bea \xi^{(3)}_{\Phi}(\k_1, \k_2, \k_3) &=& \fnl \big[P_1 P_2 + \mathrm{5 \, perms.} \big] + \mathcal{O}(\fnl^3)\ , \label{equ:tnl3}\\
\xi^{(4)}_{\Phi}(\k_1, \k_2, \k_3, \k_4)& = & 2\left (\tfrac{5}{6} \right )^2 \tnl \big[P_1P_2 P_{13} + \mathrm{23 \, perms.} \big] + \mathcal{O}(\tnl^2) \ , \label{equ:tnl4}
\eea
where we have defined $P_i \equiv P_\Phi(k_i)$ and $P_{ij} \equiv P_\Phi(|{\k}_i + {\k}_j|)$. However, unlike the single-field local ansatz, now $\tnl$ need not be equal to $\left ( \frac{6}{5} \fnl \right )^2 $. Instead, the ansatz (\ref{eqn:taunlmodel}) implies $\tnl \equiv \left ( \frac{6}{5} \fnl \right )^2 (1+\Pi)$, in agreement with the Suyama-Yamaguchi inequality, $\tnl \ge \left ( \frac{6}{5} \fnl \right )^2$~\cite{SY} (see also~\cite{Sugiyama:2011jt, Lewis:2011au, Smith:2011if, Quasi4, Kehagias:2012pd}). 
The following limits will be useful in computing the cumulants required in the barrier crossing calculation:
\begin{align}
\lim_{k_1 \to 0 } \xi^{(3)}_{\Phi}({\k}_1, {\k}_2, {\k}_3) &= 4 \fnl P_1 P_2 \label{equ:A1}\ , \\
\lim_{k_1 \to 0 }   \xi^{(4)}_{\Phi}({\k}_1,{\k}_2,{\k}_3,{\k}_4) &=   8 \left( \tfrac{5}{6} \right) ^2 \tnl \, P_1 \big[ P_2 P_3 +  P_2 P_4 +  P_3 P_4 \big] \ , \label{equ:A2} \\
\lim_{k_{12} \to 0 }   \xi^{(4)}_{\Phi}({\k}_1,{\k}_2,{\k}_3,{\k}_4) &= 16 \left( \tfrac{5}{6} \right) ^2 \tnl \, P_{12} P_1 P_3\ . \label{equ:A3} \end{align}
However, before we discuss the explicit barrier crossing result, we present an alternative derivation using the peak-background split approach.

\subsubsection{Peak-Background Split}
\label{sssec:tnl_pbs}

PBS is a heuristic procedure for predicting the large-scale clustering statistics of dark matter halos.
All fields are split into long and short modes---i.e.~the Gaussian fields in eq.~(\ref{eqn:taunlmodel}) are written as $\phi = \phi_s + \phi_\ell$ and $\psi = \psi_s + \psi_\ell$. 
The short scales ($\lesssim R_\M \lesssim 10$~Mpc$/h$) determine halo formation, while the long scales ($\gtrsim 100$ Mpc$/h$) are the ones on which we want to measure the clustering of halos. Long modes are therefore always much larger than the Lagrangian size of the halos that we consider, i.e.~$R_\M  k_{\ell} \ll 1 $.
The precise split into long and short modes isn't important for physical observables, as long as it satisfies the above constraints.

The long-wavelength modes alter the statistical properties of the small-scale fluctuations. For instance,
to lowest order, the locally measured small-scale power is
$\sigma = \bar \sigma \left[1 + 2\fnl (1+\Pi) \psi_\ell \right]$, and the locally measured halo number density is
\beq
n_\h({\x}) = \bar n_\h\left( \delta_c-\delta_\ell \, ; \bar\sigma  \left[1 + 2 \fnl (1+\Pi) \psi_\ell \right]\right)\ .
\eeq
Taylor expanding this expression, we get
\beq
\delta_\h \equiv \frac{\delta n_\h}{\bar n_\h} = b_g \delta_\ell + \beta_f(1+\Pi) \fnl \psi_\ell\ \ ,
\eeq
where
\beq
b_g \equiv \frac{\partial \ln n_\h}{\partial \delta_\ell}\quad {\rm and} \quad \beta_f \equiv 2 \frac{\partial \ln n_\h}{\partial \ln \sigma} \ . \label{equ:coefs}
\eeq
Hence, we find
\beq
P_{\rm mh} = \left( b_g + \beta_f \frac{\fnl}{\alpha(k)}\right) P_{\rm mm}\ ,
\eeq
and
\begin{align}
P_{\rm hh} 
&= \left( b_g^2 + 2 b_g \beta_f \frac{\fnl}{\alpha(k)} + \beta_f^2 \frac{\left(\tfrac{5}{6}\right)^2 \tnl}{\alpha^2(k)} \right) P_{\rm mm} \ .
\end{align}
This leads to large-scale halo stochasticity of the form
\beq
r = \left( \left(\tfrac{5}{6}\right)^2 \tnl - \fnl^2 \right) \frac{\beta_f^2}{\alpha^2(k)}\ . \label{equ:rPBS}
\eeq
As $\Pi \rightarrow 0$, this reduces to the classic $\fnl$ model, with $\tnl = \left(\frac65 \fnl \right)^2$ and hence no stochasticity.

\subsubsection{Barrier Crossing}
\label{sssec:tnl_barrier_crossing}

Next, we  show that eq.~(\ref{equ:rPBS}) can be reproduced precisely from the barrier crossing analysis of the previous section. In Appendix~\ref{sec:convergence}, we show that only the lowest-order cumulants will be significant.
Here, we calculate the relevant cumulants explicitly:
Using eq.~(\ref{equ:A1}),
we get
\beq
f_{1,2} = f_{\hat 1,2} \ \xrightarrow{k\to 0} \ 4\, \frac{\fnl}{\alpha(k)}\ . \label{equ:316}
\eeq
The order-of-magnitude estimates in Appendix~\ref{sec:convergence} suggest that this will be the dominant contribution. In particular, we expect, $f_{1,2} > f_{1,3}$. We can confirm this explicitly.
Using eq.~(\ref{equ:A2}),
we get 
\beq
f_{1,3} = f_{\hat 1,3} \ \xrightarrow{k\to 0} \ 4\, \frac{ \left( \tfrac56 \right)^2 \tnl }{ \alpha(k) } \cdot \kappa_{3}^{(\fnl = 1)}\ ,  \label{equ:318}
\eeq
where
\beq
\kappa_{3}^{(\fnl = 1)}(M) \equiv \frac{6}{\sigma_{\M}^{3}} \int \limits_{{\q}_1} \int \limits_{{\q}_2}  \alpha_\M(q_1) \alpha_\M(q_2)  \alpha_\M(q_{12})  P_{\Phi}(q_1) P_{\Phi}(q_2) \ . \label{equ:k1}
\eeq
% In~\cite{LoVerde:2011iz}, a fitting function for eq.~(\ref{equ:k1}) was presented
% \beq
% \kappa_3^{(\fnl=1)}(M) = 6.6\times 10^{-4} \cdot \left[\, 1 - 0.016 \log\left(\frac{M}{h^{-1} M_\odot}\right) \, \right] \ . 
% \eeq
% Note that eq.~(\ref{equ:318}) corresponds to an effective $\fnl$ of $\fnl^{\rm eff} = \left(\tfrac{5}{6}\right)^2 \tnl \kappa_3^{(\fnl = 1)}$. Using the current observational bound, $|\tnl| \lesssim 10^4$, gives $\fnl^{\rm eff} \lesssim {\cal O}(5)$.
% In the following, we will ignore the small window in which (\ref{equ:318}) is comparable in size to (\ref{equ:316}), and focus on the much more plausible regime, $f_{1,2} > f_{1,3}$.
Since $\kappa_3^{(\fnl=1)}$ is of order $\Delta_\Phi$, we see that the condition $f_{1,3} \ll f_{1,2}$ is equivalent to
$\fnl(1+\Pi) \Delta_\Phi \ll 1$.  This latter condition is always satisfied if all fields are weakly coupled.\footnote{In
more detail, to show that $\fnl (1+\Pi) \Delta_\Phi \ll 1$, we
argue as follows.  Assuming that the field $\psi$ is not strongly coupled, the dimensionless non-Gaussianity parameter
$\fnl^{(\psi)} \Delta_\psi = \fnl (1+\Pi)^{3/2} \Delta_\Phi$ must be $\lsim 1$.  Therefore
\beq
\fnl (1+\Pi) \Delta_\Phi = \Big[ \fnl \Delta_\Phi \Big]^{1/3} \cdot \Big[ \fnl (1+\Pi)^{3/2} \Delta_\Phi \Big]^{2/3} \lsim [10^{-3}]^{1/3} \cdot [1] = 10^{-1}\ ,
\eeq
where the bound on the first factor is the current observational bound $\fnl \lsim 10^2$.}

Finally, using eq.~(\ref{equ:A3}),
we get 
\beq
f_{2,2} \xrightarrow{k\to 0} 16\, \frac{ \left( \tfrac56 \right)^2 \tnl}{\alpha^2(k)}\ .
\eeq
Substituting the above into eq.~(\ref{equ:master}) gives
\beq
r = \left( \left(\tfrac{5}{6}\right)^2 \tnl - \fnl^2 \right) \frac{\beta_f^2}{\alpha^2(k)}\ ,
\eeq
where we have used the relation $\beta_f = 4 \beta_2 = 2(\nu_c-1)$,
which can be derived by evaluating the derivative $\beta_f = 2 \hskip 1pt \partial\ln n_\h/\partial\ln\sigma$ in the barrier crossing model~\cite{Slosar:2008hx}.
Comparing with eq.~(\ref{equ:rPBS}), we find that barrier crossing and peak-background split give consistent answers.

\subsection{$g_{\rm NL}$ Cosmology} 
\label{sec:gnl}

As our next example, we consider a cubic form of local non-Gaussianity.\footnote{We should say from the outset 
that the large-scale stochasticity in the $\gnl$ model will be too small to be observationally relevant.
Although the non-stochastic and stochastic contributions to $P_{\h\h}(k)$ will turn out to be parametrically identical
($\sim \gnl^2 \Delta_\Phi^2 P_\Phi(k)$), the non-stochastic contribution is typically larger by a constant factor $\approx 10^4$.
Nevertheless, the $\gnl$ example provides an interesting check of our formalism.}
In this case, the non-Gaussian potential is parametrized by the expansion
\beq
\Phi = \phi + \gnl \left(\phi^3 - 3 \langle \phi^2 \rangle \phi \right)\ .
\eeq
The power spectrum of the non-Gaussian field is
\beq
P_\Phi(k) = P_\phi(k) + \gnl^2 P_{\phi^3}(k)\ ,
\eeq
where
\beq
P_{\phi^3}(k) \equiv 6 \int_{{\q_1}} \int_{{\q_2}} P_\phi(q_1) P_\phi(q_2) P_\phi(|{\k}-{\q_1}-{\q_2}|)\ .
\eeq
We note that for scale-invariant initial conditions, $(k^3/2\pi^2)P_\phi(k) = \Delta_\phi^2 $, the power spectrum $P_{\phi^3}$ is infrared divergent. If the IR divergence is regulated by putting the fields in a finite box with length $L$, then the power spectrum diverges as 
\beq
P_{\phi^3}(k) \, \sim\, 18 \hskip 1pt \Delta_{\phi}^4 \ln^2(k L) P_\phi (k) \ .
\eeq
On large scales, the matter power spectrum therefore is
\beq
P_{\m \m}(k) \simeq \alpha^2(k) P_\Phi(k) = P_{g}(k) \left( 1+ 18 \hskip 1pt \gnl^2\Delta_{\phi}^4 \ln^2(k L) \right) \ ,
\eeq
where we defined $P_{g}(k) \equiv \alpha^2(k) P_\phi(k)$.
Current observational constraints imply that $|\gnl \Delta_\phi^2| \ll 1$.
To obtain answers to zeroth or first order in $\gnl \Delta_\phi^2$, it suffices to set  $P_{\m \m} \simeq P_{g}$.
 
For the barrier crossing analysis,
we require the following higher-order correlation functions
\bea 
% \xi^{(3)}_{\Phi}(\k_1, \k_2, \k_3) & = &{\cal O}(\gnl^3)\ , \label{equ:A1g} \\ 
\xi^{(4)[{\rm tree}]}_{\Phi}(\k_1, \k_2, \k_3, \k_4) & = & \gnl \big[P_1 P_2 P_3 + \mathrm{23 \, perms.} \big] + \mathcal{O}(\gnl^2)\ , \label{equ:A2g} \\
\xi^{(4)[{\rm loop}]}_{\Phi}({\k}_1,{\k}_2,{\k}_3,{\k}_4) &= & 9 \hskip 1pt\gnl^2  \big[P_{1} P_2 P_{\phi^2}({k}_{13}) + 11 \, {\rm perms.} \big]\ , \label{equ:A3g} \\
\xi^{(6)}_{\Phi}({\k}_1,{\k}_2,{\k}_3,{\k}_4, {\k}_5, {\k}_6) &= & 36\hskip 1pt \gnl^2 \big[ P_1 P_2 P_3 P_4 P_{125} + 89\, {\rm perms.} \big]\ , \label{equ:A4g}
 \eea
where $k_{ij}=|\k_i+\k_j|$, $P_i = P_\phi(k_i)$, $P_{ijk} = P_\phi(|\k_i+\k_j+\k_k|)$,  and
\beq
P_{\phi^2}(k) \equiv 2 \int_{{\q}} P_\phi(q) P_\phi(|{\k}-{\q}|) \ \sim\ 4 \hskip 1pt \Delta_\phi^2 \ln(kL) P_\phi(k)\ .
\eeq
Note that odd-point correlation functions $\xi^{(2N+1)}_{\Phi}$ are zero due to the $\Phi \rightarrow -\Phi$ symmetry.
Next, we will derive the stochastic halo bias both in peak-background split and in barrier crossing.

\subsubsection{Peak-Background Split}
\label{sssec:gnl_pbs}

The PBS analysis proceeds as before.
Splitting the Gaussian potential into long and short modes, $\phi = \phi_\ell + \phi_s$, we find that the locally measured small-scale power is
$\sigma = \bar \sigma \left[1+ 3 \gnl \left( \phi_\ell^2 - \langle \phi_\ell^2 \rangle \right) \right]$. 
Moreover, the locally measured value of $\fnl$ is
$\fnl^{\rm eff} = 3 \gnl \phi_\ell $ \cite{Smith:2011ub}. 
The %locally measured 
halo number density therefore is
\beq
n_\h({\x}) = \bar n_\h\left( \delta_c-\delta_\ell \, ; \bar\sigma \left[1+ 3 \gnl \left( \phi_\ell^2 - \langle \phi_\ell^2 \rangle \right) \right] ; \fnl^{\rm eff}\right)\ ,
\eeq
where $\delta_\ell \simeq \alpha(k_\ell) \phi_\ell$.  Taylor expanding this expression, we find
\beq
\delta_\h  = b_g \delta_\ell + \tfrac{3}{2}\beta_f \gnl  \left( \phi_\ell^2 - \langle \phi_\ell^2 \rangle \right) + \beta_g \gnl \phi_\ell\ ,
\eeq
where $b_g$ and $\beta_f$ are the same as in (\ref{equ:coefs}), and
\beq
 \beta_g \equiv 3 \frac{\partial \ln n_\h}{\partial \fnl}\ .
\eeq
It follows that
\beq
P_{\rm mh} = b_g P_{\rm mm} + \beta_g \gnl P_{{\rm m} \phi} = \left( b_g + \beta_g \frac{\gnl}{\alpha(k)}\right) P_{\rm mm}\ ,
\eeq
and 
\beq
P_{\h \h} = \left( b_g + \beta_g \frac{\gnl}{\alpha(k)} \right)^2 P_{\rm mm} + \frac94 \hskip 1pt \beta_f^2 \gnl^2 P_{\phi^2} \ .
\eeq
This implies a large-scale halo stochasticity of the form
\beq
r = \frac{9}{4}\hskip 1pt \beta_f^2 \gnl^2 \frac{P_{\phi^2}}{P_{\rm mm}} \ . \label{equ:PBSx}
\eeq

\subsubsection{Barrier Crossing}
\label{sssec:gnl_barrier_crossing}

We now show that the same result is obtained from barrier crossing.
In Appendix~\ref{sec:convergence}, we argue that only the first few cumulants need to be taken into account. It is straightforward to compute them explicitly.
From eqs.~(\ref{equ:A2g}) and (\ref{equ:A4g}),
we get
\begin{align}
f_{1, 3}(k) &\ \xrightarrow{k\to 0} \ \frac{3 \gnl}{\alpha(k)} \, \kappa_{3}^{(f_{\rm NL} = 1)}\qquad {\rm and} \qquad 
f_{3,3}(k) = [f_{1, 3}(k)]^2\ .
\end{align}
This only contributes to the non-stochastic bias.
However, since $f_{\hat 1,2} = 0$, stochastic bias arises from $f_{2,2}$.
First, we note that the tree-level four-point function (\ref{equ:A2g}) leads to a very small and scale-independent contribution to $f_{2,2}$\,:
\begin{align}
% f_{2, 2}^{[{\rm tree}]}(k) &\ \xrightarrow{k\to 0} \gnl \cdot \frac{24}{\sigma_{\M}^2} \frac{\int_\q \alpha_{\M}^2(q)[P_\phi(q)]^2}{P_{\m \m}(k)}\ .
f_{2,2}^{[{\rm tree}]}(k,M,\bar M) &\ \xrightarrow{k\to 0} \ \frac{12 \hskip 1pt \gnl}{P_{\m\m}(k)} \left(
  \frac{1}{\sigma_{\M}^2} \int_{\q} \alpha_{\M}^2(q) P_\phi^2(q) +
  \frac{1}{\sigma_{\bar\M}^2} \int_{\q} \alpha_{\bar\M}^2(q) P_\phi^2(q) \right)\ .
\end{align}
Plugging into eq.~(\ref{equ:master}) and noting that $\beta_2 = \frac{1}{4}\beta_f$ and $\tilde\beta_2 = \frac{1}{2}$,
we get a small scale-dependent contribution to the large-scale stochastic bias
\beq
r_{[{\rm tree}]} = \frac{3}{2} \frac{\gnl}{P_{\m\m}(k)} \left( \beta_f^2 + 2 \beta_f \right) 
  \left( \frac{1}{\sigma_{\M}^2} \int_{\q} \alpha_{\M}^2(q) P_\phi^2(q) \right)  \ . \label{eq:gnl_rtree}
\eeq
% This small contribution to the stochasticity is only reproduced in PBS if we expand $n_\h$ to second order in $\delta_{\l}$ (see \S\ref{sec:Alocal}).
% In practice, this contribution to the large-scale stochasticity can't be used as a probe of initial conditions, since it is degenerate with other sources of stochasticity such as second-order Gaussian bias.
In practice, this contribution to the large-scale stochasticity can't be used as a probe of initial conditions, since a
contribution to $r$ with $r \propto 1/k$ % scale dependence 
(or equivalently a contribution to $P_{\h\h}(k)$ which approaches
a constant as $k\rightarrow 0$) is degenerate with other sources of stochasticity such as second-order Gaussian bias.
%\footnote{\kms{I think we went back and forth on the preceding sentence, but I like the more quantitative version which makes it clear that the $r_{[{\rm tree}]} \propto (1/k)$ scale dependence is the reason we can't use $r_{[{\rm tree}]}$ to probe initial conditions, whereas  $r_{[{\rm loop}]} \propto (1/k^4)$ is OK.}}
Finally, the one-loop four-point function (\ref{equ:A3g}) leads to the following contribution to $f_{2,2}$\,:
\begin{align}
f_{2, 2}^{[{\rm loop}]}(k) &\ \xrightarrow{k\to 0} \ 36 \, \gnl^2 \cdot \frac{P_{\phi^2}(k)}{P_{\m \m}(k)}\ .
\end{align}
The corresponding stochasticity parameter is
\beq
r_{[{\rm loop}]} = \frac{9}{4}\hskip 1pt \beta_f^2 \gnl^2 \frac{P_{\phi^2}(k)}{P_{\rm mm}(k)} \propto \frac{1}{k^4}\ ,  \label{eq:gnl_rloop}
\eeq
in agreement with the PBS predictions (\ref{equ:PBSx}).

%\newpage
\subsection{Quasi-Single-Field Inflation}

Our last example is quasi-single field inflation (QSFI) \cite{Chen:2009zp}.  These models involve extra massive scalar degrees of freedom during inflation.  In the simplest examples, a single scalar field $\sigma$ of mass\footnote{We note that extra scalars with masses close to the Hubble scale $H$ are a natural prediction of supersymmetric theories of inflation (see~\cite{Baumann:2011nk} for further discussion).} $m^2 \leq \tfrac{9}{4} H^2$ mixes with the fluctuation of the inflaton\footnote{Recall that $\delta \phi$ in spatially flat gauge is proportional to the curvature perturbation, $\zeta \equiv \frac{H}{\dot \phi} \delta \phi$.} $\delta \phi$.
The mixing communicates non-Gaussianity from the hidden (isocurvature) sector to the observable (adiabatic) sector. As we now show, it also leads to a significant stochasticity in the halo bias.

\newpage
\subsubsection{Boosted Four-Point Function}

%\kms{I didn't equation check this subsection since DG or DB can probably do it very quickly} \db{Will do.}
Again, we need the squeezed and collapsed limits of the primordial correlation functions\footnote{See \cite{Baumann:2011nk} for an intuitive explanation of the scalings in eqs.~(\ref{equ:A1a}) and (\ref{equ:A3a}).}~\cite{Chen:2009zp, Baumann:2011nk,Quasi4}: 
\begin{align}
\lim_{k_1 \to 0 } \xi^{(3)}_{\Phi}({\k}_1, {\k}_2, {\k}_3) &= 4 \fnl \left( \frac{k_1}{k_2}\right)^{\Delta} P_1 P_2 \label{equ:A1a}\ , \\
\lim_{k_{12} \to 0 }   \xi^{(4)}_{\Phi}({\k}_1,{\k}_2,{\k}_3,{\k}_4) &= \  16 \left( 
\tfrac{5}{6}\right)^2 \tnl   \left( \frac{k_{12}^2}{k_{1} k_{3}}\right)^{\Delta} P_1 P_3 P_{12} \ , \ \  \label{equ:A3a} \end{align}
where we defined the parameter 
\beq
\Delta \equiv \frac{3}{2} - \sqrt{\frac{9}{4} - \frac{m^2}{H^2} }\ . 
\eeq
The non-trivial momentum scaling of eqs.~(\ref{equ:A1a}) and (\ref{equ:A3a}) is a remarkable signature of extra Hubble mass scalars during inflation~\cite{Chen:2009zp,Baumann:2011nk, Sefusatti:2012ye, Norena:2012yi}.
Moreover, if the mixing between $\sigma$ and $\phi$ (or $\zeta$) is parametrized by a small dimensionless number $\varepsilon < 1$, then 
\beq
\tnl \sim  \varepsilon^{-2} \, (\tfrac{6}{5}\fnl)^2 \ >\  (\tfrac{6}{5}\fnl)^2 \ .
\eeq
The enhancement of $\tnl$ arises because the trispectrum is generated by the exchange of the $\sigma$-field which is only weakly coupled to $\zeta$. 
The size of the four-point function $\langle \zeta^4 \rangle$ can be estimated
from the square of the three-point function $ \langle \zeta ^2 \sigma \rangle$ at horizon crossing,
\beq
\langle\zeta^4 \rangle \sim \langle \zeta ^2 \sigma \rangle^2 \sim \varepsilon^{-2}\hskip 2pt  \frac{\langle \zeta^3 \rangle^2}{\langle \zeta^2 \rangle}\ .
\eeq 
 The boost of $\tnl$ is the result of the small correlation between the curvature fluctuation and the massive field, $\varepsilon \ll 1$. The precise dependence of $\fnl$ and $\tnl$ on the fundamental parameters of the QSFI Lagrangian can be found in \cite{Quasi4}.

\subsubsection{Barrier Crossing}

In QSFI, the higher-order $N$-point functions are suppressed by factors of the power spectrum, just as in our previous examples.  The dominant contributions to the large-scale structure signal therefore arise from the squeezed limit of the three-point function and the collapsed limit of the four-point function.  The relevant cumulants are
\begin{align}
\kappa_{\hat{1},2} (k) &\, \xrightarrow{k\to 0}\, \frac{4 \fnl}{\hat \sigma} (k R_\M)^{ \Delta} \, \frac{P_{\rm mm} (k)}{\alpha(k)} \,\frac{\Sigma^2_{\M}(\Delta) }{\sigma_{\M}^2 }\ ,
\end{align}
and 
\begin{align}
\kappa_{2,2} (k) &\, \xrightarrow{k\to 0}\,  16 \left( 
\tfrac{5}{6}\right)^2 \tnl (k^2 R_\M R_{\bar \M})^{\Delta} \, \frac{P_{\rm mm}(k)}{\alpha^2(k)} \, \frac{\Sigma^2_{\M}(\Delta)}{\sigma_{\M}^2 }\frac{\Sigma^2_{\bar \M}(\Delta)}{\sigma_{\bar \M}^2 } \ .
\end{align}
Here, we have defined %\ssf{Missing a factor of $1/(2 \pi)^3$?}
\beq
\Sigma^2_\M(\Delta) \equiv \int \frac{\d^3 k_s}{(2\pi)^3} \, W^2_{\M}(k_s)(k_s R_\M) ^{- \Delta } P_{\rm mm}(k_s) \ , \label{equ:Sigma}
\eeq
where the integration variable, $k_s$, is one of the short momenta and $R_\M$ is the smoothing scale defined by eq.~(\ref{eqn:tophat}). By definition, $\Sigma_\M(0) = \sigma_\M$.  
In the limit $\Delta \to 0$, we recover the results of the $\tnl$~model.  Therefore, we find
\beq
f_{\hat 1, 2} = 4 \fnl  \, \frac{(k R_\M)^{ \Delta}}{\alpha(k)} \,\frac{\Sigma^2_{\M}(\Delta) }{\sigma_{\M}^2 } \qquad {\rm and} \qquad f_{2,2}= \frac{\tnl}{(\tfrac{6}{5} \fnl)^2} \, f_{\hat 1, 2}(M) f_{\hat 1, 2}(\bar M) \ . \label{equ:fs}
\eeq

\begin{figure}[h!]
   \centering
       \includegraphics[height=6.7cm]{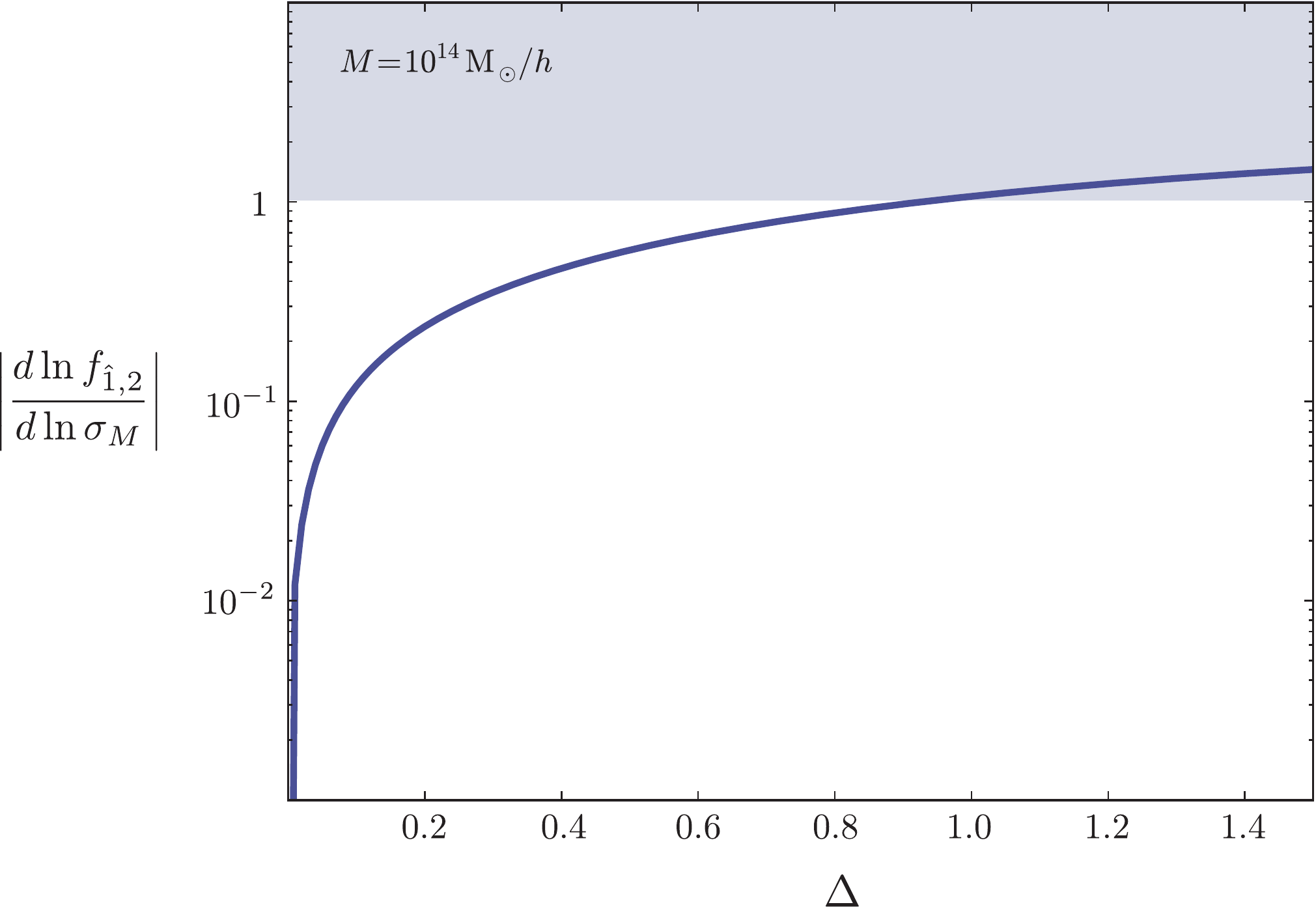}
   \caption{Numerical evaluation of eqs.~(\ref{equ:Sigma}) and (\ref{equ:fs}). For $\Delta \gtrsim 1.0$, the cumulant $f_{\hat 1,2}$ depends significantly on the halo mass scale~$M$. This is in contrast to local non-Gaussianity, which corresponds to the limit $\Delta \to 0$.}
%\kms{Suggest removing $k = 0.01$ $h$/Mpc from the figure panel, since the quantity on the $y$-axis is independent of $k$}}
  \label{fig:QSFI}
\end{figure}

\noindent
To obtain the large-scale stochasticity, we substitute the cumulants into eq.~(\ref{equ:master}),
\begin{align}
r \ =\ \left( \beta_2 + \tilde\beta_2 \frac{\partial}{\partial\ln\sigma_{\M}} \right)
     \left( \beta_2 + \tilde\beta_2 \frac{\partial}{\partial\ln\sigma_{\bar\M}} \right)
     f_{2,2} \, -\,  \left[ \left( \beta_2 + \tilde\beta_2 \frac{\partial}{\partial\ln\sigma_{\M}} \right)
     f_{\hat 1,2} \right]^2\ .
\end{align}
Because the cumulants depend explicitly on $R_\M$, we 
 have to be concerned that the derivatives with respect to $\sigma_\M$ may this time not be negligible. 
Indeed, numerical evaluation of the integral shows significant $\sigma_\M$ dependence of $f_{\hat 1, 2}$ (see fig.~\ref{fig:QSFI}).
Keeping the derivative terms, we get 
\begin{align}
r &\ =\ \left( \left ( \tfrac{5}{6} \right )^2 \tnl -   \fnl ^2 \right) \frac{k^{2\Delta}}{\alpha^2(k)} 
      \left[ \left( \beta_f + 2\, \frac{d}{d\ln\sigma_{\M}} \right) \frac{R_{\M}^{\Delta}\hskip 1pt \Sigma_{\M}^2(\Delta)}{\sigma_{\M}^2} \right]^2 \nonumber \\
  &\ \propto \ \frac{\left ( \tfrac{5}{6} \right )^2 \tnl -   \fnl ^2 }{k^{4 - 2 \Delta}} \ . \label{equ:rQSFI}
\end{align}
The characteristic momentum scaling of eq.~(\ref{equ:rQSFI}) and the natural boost of $\tnl$ makes halo stochasticity an interesting probe of quasi-single-field inflation.

\newpage
\section{Conclusions}
\label{sec:conclusions}

What was the number of light degrees of freedom during inflation? 
And, what were their interactions?
The great virtue of primordial non-Gaussianity is that it is sensitive to these basic questions about the physics of inflation.
In particular, it is well-known that the {squeezed} limit of the primordial {three-point} function, 
\beq
\lim_{k_1 \to 0} \langle \Phi_{\k_1} \Phi_{\k_2} \Phi_{\k_3} \rangle\ ,
\eeq
can only be large if more than one light field was dynamically relevant during inflation~\cite{Maldacena:2002vr,Creminelli:2004yq}. Remarkably, this statement is independent of the details of the Lagrangian for the inflaton field and its initial conditions.
Measurements of the squeezed limit therefore have the potential to rule out all models of single-field inflation \cite{Maldacena:2002vr,Creminelli:2004yq}. 
Moreover, the precise scaling in the squeezed limit is sensitive to the details of the mass spectrum~\cite{Chen:2009zp, Quasi4}, allowing a test of extra Hubble mass fields, such as those generically expected in supersymmetric theories~\cite{Baumann:2011nk}.
Having a large three-point function in the squeezed limit modulates the two-point function of halos and therefore leads to scale-dependent bias~\cite{Dalal:2007cu}. In the future, this effect may well be our most sensitive probe of the squeezed limit. 

In this paper, we have discussed a {\it stochastic} form of scale-dependent halo bias.
This effect arises if the {collapsed} limit of the primordial {four-point} function,
\beq
\lim_{k_{12} \to 0} \langle \Phi_{\k_1} \Phi_{\k_2} \Phi_{\k_3}  \Phi_{\k_4} \rangle_{\rm c} \ ,
\eeq
is larger than the square of the squeezed limit of the three-point function.
More generally, stochastic bias arises whenever a suitable collapsed limit of an $(M+N)$-point function is larger than the
product of the associated squeezed $(M+1)$-point and $(N+1)$-point functions, where $M,N\ge 2$.  
The key tool for obtaining this result, and a main result of this paper, is a pair
of formulas, eqs.~(\ref{equ:PmhFinal}) and~(\ref{equ:PhhFinal}), for the matter-halo and halo-halo power spectra in a general
non-Gaussian model parametrized by the $N$-point functions of the primordial potential.

In non-Gaussian models which generate significant stochastic halo bias, the results of this paper
are important even at a qualitative level.
As a concrete example, it should be possible to measure $\fnl$ and $\tnl$ independently using stochastic bias.
This can be done either by measuring multiple tracer populations and directly estimating large-scale stochasticity
(which has the advantage of eliminating sample variance), or from a single tracer population by measuring $P_{\h\h}(k)$
and using the functional form
\beq
P_{\h\h}(k) = b_g^2 \left( 1 + \fnl \frac{2 \delta_c}{\alpha(k)} + \tnl \frac{(\frac{5}{6})^2 \delta_c^2}{\alpha^2(k)} \right)
\eeq
to fit for $b_g$, $\fnl$ and $\tnl$ independently.
Recently, ref.~\cite{Biagetti:2012xy} showed that if only non-stochastic bias is considered, the leading contribution from $\tnl$ is small
(in our language, this corresponds to the ${\mathcal O}(\tnl)$ contribution to $\kappa_{1,3}$) and it is difficult to separate $\fnl$ and $\tnl$,
so stochastic bias has an important qualitative effect.
As another example, in quasi-single field inflation, the stochastic bias is larger than the non-stochastic bias
by a large factor (parametrically $\varepsilon^{-2}$), leading to a similarly large enhancement in signal-to-noise when stochastic bias
is considered.
We defer quantitative forecasts incorporating stochastic bias for future work.

%It is easy to show that 
In general, there is no stochastic bias if only a single field (which may or may not be the inflaton) 
generates the primordial curvature perturbation and its non-Gaussianity~\cite{Suyama:2010uj}. %\kms{This statement matches my intuition,but how does one show this?  It doesn't seem obvious to me} \db{I have in mind \cite{Suyama:2010uj}, but I wouldn't call this proof, so I would be happy to weaken the statement.} \dg{ There are exceptions in the non-Bunch-Davies vacuum.  These examples will only produce stochastic bias over a finite range of momentum so the statement here is analogous to the one about the squeezed limit.}  
Measuring stochastic halo bias would therefore teach us about the effective number of degrees of freedom that generated the primordial fluctuations and its higher-order correlations. In particular, stochasticity is sensitive to what we may call ``hidden sector non-Gaussianity", i.e. situations in which two fields generate the curvature perturbation, but only one (hidden) field is responsible for its non-Gaussianity.
In this paper, we have derived this effect for general non-Gaussian initial conditions. We have also applied our formalism to a number of explicit examples, such as curvaton models \cite{Tseliakhovich:2010kf} and quasi-single field inflation~\cite{Chen:2009zp}.
%In conclusion, w
We have shown that halo bias, in principle, gives us information about the soft limits of both the primordial three-point function and the four-point function. It is therefore a valuable tool in the quest to uncover the physics that created the initial perturbations.

\acknowledgments
We thank Valentin Assassi, Eugene Lim, Marilena LoVerde, Marcel Schmittfull, David Spergel and Matias Zaldarriaga for helpful discussions.
D.B.~gratefully acknowledges support from a Starting Grant of the European Research Council (ERC STG grant 279617). S.F.~acknowledges support from a fellowship at the Department of Astrophysical Sciences of Princeton University.
The research of D.G.~is supported by the DOE under grant number DE-FG02-90ER40542 and the Martin A.~and Helen Chooljian Membership at the Institute for Advanced Study.
K.M.S.~was supported by a Lyman Spitzer fellowship in the Department of Astrophysical Sciences at Princeton University.
Research at Perimeter Institute is supported by the Government of Canada
through Industry Canada and by the Province of Ontario through the Ministry of Research \& Innovation.

\newpage
\appendix
\section{Convergence of the Edgeworth Expansion}
\label{sec:convergence}

%\db{fix intro}

In this appendix, we discuss the convergence properties of the Edgeworth expansion for local non-Gaussianity.
In particular, we will estimate the relative size of the cumulants $\kappa_{n,m}(k,M)$ for general $n$ and $m > 0$ in the large-scale limit $k \to 0$.  
For further discussion see e.g.~\cite{LoVerde:2007ri, Shandera:2008ai, Barnaby:2011pe}.
The results in this appendix are used in the main text in several places: to justify the approximation that non-linear terms in the Edgeworth
expansion are negligible in eqs.~(\ref{eq:xi_mh_linear}) and~(\ref{equ:xi3}), and to justify keeping only certain cumulants in the $\tnl$ model
(\S\ref{sssec:tnl_barrier_crossing}) and the $\gnl$ model (\S\ref{sssec:gnl_barrier_crossing}).

\subsection{$\tau_{\rm NL}$ Cosmology}

We first consider the $\tnl$ model of \S\ref{sec:tnl}.

\vskip 4pt
\noindent
{\it Linear terms.}---The leading contribution in the $k\to 0$ limit arises from the following contribution to the connected correlation function
\beq \label{eqn:contraction}
\contraction{\kappa_{n,m} \, \simeq\, {\cal A}_{n,m} \int \langle }{\psi}{_1(}{\psi}
\contraction{\kappa_{n,m} \, \simeq\, {\cal A}_{n,m} \int \langle \psi_1(\psi }{\psi}{)_2}{}
\contraction{\kappa_{n,m} \, \simeq\, {\cal A}_{n,m} \int \langle \psi_1(\psi \psi)_2 \cdots }{{\color{white} \psi}}{ (}{\psi}
\contraction{\kappa_{n,m} \, \simeq\, {\cal A}_{n,m} \int \langle \psi_1(\psi \psi)_2 \cdots {\color{white} \psi}(\psi }{\psi}{)_n\ |\ (}{\psi}
\contraction{\kappa_{n,m} \, \simeq\, {\cal A}_{n,m} \int \langle \psi_1(\psi \psi)_2 \cdots {\color{white} \psi}(\psi \psi)_n\ |\ (\psi }{\psi}{)_{n+1}}{ }
\contraction{\kappa_{n,m} \, \simeq\, {\cal A}_{n,m} \int \langle \psi_1(\psi \psi)_2 \cdots {\color{white} \psi}(\psi \psi)_n\ |\ (\psi \psi)_{n+1} \ldots }{{\color{white} \psi}}{(}{\psi}
\contraction{\kappa_{n,m} \, \simeq\, {\cal A}_{n,m} \int \langle \psi_1(\psi \psi)_2 \cdots {\color{white} \psi}(\psi \psi)_n\ |\ (\psi \psi)_{n+1} \ldots {\color{white} \psi}(\psi }{\psi}{)_{n+m-1}} {\psi}
\kappa_{n,m} \, \simeq\, {\cal A}_{n,m} \int \langle \psi_1(\psi \psi)_2 \cdots {\color{white} \psi}(\psi \psi)_n\ |\ (\psi \psi)_{n+1} \ldots {\color{white} \psi} (\psi \psi)_{n+m-1} \psi_{n+m} \rangle_{\rm c}' \ \d  \bf K \ ,  \eeq
where $\d {\bf{K}}  \equiv \prod_i \frac{\d^3 \k_i}{(2\pi)^3}\, \alpha_\M(k_i)$ and $(\psi \psi)_i$ denotes an auto-convolution evaluated at $\k_i$. 
The prime on the correlation function denotes that we have dropped an overall momentum conserving delta-function.
 The amplitude of the cumulant is given by
\beq\label{eqn:contractionamp}
{\cal A}_{n,m} \equiv c_{n,m} (1+\Pi)^{2(n+m-2)} \fnl^{n+m-2} \ , \qquad {\rm where} \qquad c_{n,m} \equiv  \frac{n! m! \, 2^{n+m-2}}{\sigma_{\M}^n \sigma_{\bar \M}^m} \ .
\eeq
We arrived at eq.~(\ref{eqn:contraction}) by using the definition of $\Phi$ in eq.~(\ref{eqn:taunlmodel}) and expanding out terms to produce a connected correlation function. The numerical factor $c_{n,m}$ in the amplitude (\ref{eqn:contractionamp}) arises from the sum over equivalent contractions of the fields.  The vertical line in (\ref{eqn:contraction}) separates the first $n$ terms from the last~$m$. Each contraction gives a factor of $P_{\psi}$, and the contraction crossing the vertical line carries momentum $k$, giving a factor of $P_{\psi}(k)$ that can be taken out of the integral.  The power spectrum $P_{\psi}(k)$ diverges as $k \to 0$ and gives the largest\footnote{Subleading contributions arise when both linear $\psi$ terms appear on the same side. %(e.g.~$\psi_{1} \leftrightarrow (\psi \psi)_{n+1}$) \dg{Is this clear?}.  
In such cases, two contractions cross the vertical line, and the resulting cumulant is finite in the $k \to 0$ limit.} contribution to $\kappa_{n,m}$. 
The remaining integral over $\d {\bf K}$ will typically be dominated by the non-linear scale $k_{\rm nl}$, where $k_{\rm nl}^3 P_{\m \m}(k_{\rm nl}) \sim \alpha^2_\M(k_{\rm nl}) \Delta_\Phi^2 \sim 1$.  Therefore, we may estimate the integral using $\alpha_\M \sim \Delta_\Phi^{-1}$, to get
\begin{align} 
\kappa_{1,m} &\ \simeq \ c_{1,m}\, (1+\Pi)^{m-2} \  \fnl^{m-1} \ \Delta_{\Phi}^{m-2} \cdot \frac{P_{\m \m}(k)}{\alpha(k)} \ \, \qquad \qquad \ \ \ \, \mathrm{for \ \ } m >1 \ ,  \label{eq:fnl_schematic2} \\
 \kappa_{n,m} &\ \simeq \ c_{n,m}\,  (1+\Pi)^{n+m-3}\  \fnl^{n+m-2} \ \Delta_{\Phi}^{n + m -4} \cdot \frac{P_{\m \m}(k)}{\alpha^2(k)} \ \ \ \ \ \ \ \mathrm{for \ \ } n \ {\rm and} \ m >1 \ . \label{eq:fnl_schematic}
\end{align}
The factor $n!m!$ appearing in $c_{n,m}$ is canceled explicitly in the Edgeworth expansion (\ref{equ:edge1}), and
as shown in \S\ref{sssec:tnl_barrier_crossing}, the condition $\fnl (1+\Pi) \Delta_\Phi \ll 1$ is always satisfied.
This implies that higher-order cumulants are subdominant relative to lower-order ones,
and hence the only terms we have to keep in the $\tnl$ model are $\kappa_{1,1}$, $\kappa_{1,2} = \kappa_{2,1}$ and $\kappa_{2,2}$.

\vskip 4pt
\noindent
{\it Non-linear terms.}---When expanding the exponential in the Edgeworth expansion (\ref{equ:edge1}) we also encounter non-linear terms such as $\kappa^P_{n,m}({\x})$.  First, we will show that, for $n$ and/or $m > 1$, these terms are suppressed by the near-Gaussianity of the primordial perturbations.  We distinguish two cases: 
%i) $n>1$ and $m>1$, and ii) $n=1$ and $m>1$. 
\begin{itemize}
\item When $n>1$ {and} $m >1$, we take powers of the contributions in (\ref{eq:fnl_schematic}), to find    
\begin{align}
\kappa_{n,m}^P &\ \sim\ \Big[c_{n,m} \, (1+\Pi)^{n+m-3}\,  \fnl^{n+m-2} \, \Delta_{\Phi}^{n+m-4}\, \Big]^P \cdot \frac{P_{\m \m}(k)}{\alpha^2(k)} \cdot \Delta_\Phi^{2(P-1)} \ln^{P-1}(k L) \ ,
\end{align}
where $L$ is an infrared cutoff. This can be written as
\begin{align}
\kappa_{n,m}^P &\ \sim\ \kappa_{n,m} \cdot c_{n,m}^P \big[ \fnl (1+\Pi) \Delta_{\Phi} \big]^{(P-1)(n+m -2)} (1+\Pi)^{-P}\ln^{P-1}(k L) \ .
\end{align}
Using $ \fnl (1+\Pi)  \Delta_\Phi \ll 1$ and $(1+\Pi)> 1$, we see that $\kappa_{n,m}^P$ is suppressed relative to $\kappa_{n,m}$ for $n,m > 1$.  

\item When $n=1$ and $m>1$, the situation is slightly different.  If we take higher powers of the results in (\ref{eq:fnl_schematic2}), we find for $P>1$,
\begin{align}
\kappa^P_{1,m} &\ \sim\ c_{1,m}^P \left[(1+\Pi)^{m-2} \,  \fnl^{m-1} \, \Delta_{\Phi}^{m-2} \right]^P \Delta_\Phi^{P-1} P_{\m\m} (k_{\rm nl})\nonumber \\
&\ \sim\ \kappa_{1,m}\cdot c_{1,m}^{P-1} \big[ \fnl (1+\Pi) \Delta_{\Phi} \big]^{(P-1) (m-1)} (1+\Pi)^{-P} \alpha(k) \cdot \frac{P_{\m \m}(k_{\rm nl})}{P_{\m \m}(k)} \label{equ:koneX}\ .
\end{align}
%\dg{I changed the above formula.  Please check.} 
Again, as we increase the power $P$, the contribution is suppressed.  However, there is a clear difference between $P=1$ and $P>1$.  Nevertheless, in the limit $k \to 0$, $[P_{\m \m}(k) / \alpha(k)]^{-1} \propto k$ so that these contributions vanish relative to $\kappa_{1,m}$.
\end{itemize}
Next, we consider products of the Gaussian piece, $\kappa_{1,1}^P$. %As we take larger powers, 
We find for~$P> 1$,
\begin{align}
\kappa_{1,1}^P &\ \sim\ P_{\m \m}^P(k_{\rm eq}) \cdot (k_{\rm eq})^{3P-3} \nonumber \\
&\ \sim\ \kappa_{1,1}\, \frac{P_{\m \m}(k_{\rm eq})}{P_{\m\m}(k)} \, \Delta_\m^{P-1}(k_{\rm eq})\label{equ:koneone} \ .
\end{align}
Here, $\kappa_{1,1}$ receives its largest contribution from the peak of the linear matter power spectrum $\Delta_\m^{2}(k) = k^3 \hat P_\m(k)$ which occurs at $k = k_{\rm eq}$, the scale set by matter-radiation equality.  Because $\Delta_\m(k_{\rm eq}) < 1$ at that scale, the modes are still linear and higher powers of $\kappa_{1,1}$ will be suppressed.  
%\db{I don't understand the next sentence:}
However, in the limit $k \to 0$, $\kappa_{1,1}$ vanishes, while $\kappa_{1,1}^P$ is finite for $P>1$. 
This gives a small constant contribution to the halo power spectrum $P_{\h \h}$ which is a free parameter in practice (we discussed this
in the context of the $\gnl$ model in~\S\ref{sssec:gnl_barrier_crossing}).
%Therefore, we cannot neglect all non-linear terms in $\kappa_{1,1}$.
%\dg{Shouldn't this worry us?  These contribute constant terms to the halo-halo correlation function as $k\to 0$ and should also contribute scale-dependent stochastic bias.  These would contribute in the gaussian case as well, so maybe it could explain the issue with the simulations.}.
%\ssf{These are (small) constant contributions to the gaussian $P_{hh}$, in addition to the $1/n$ Poisson noise term (can be seen as corrections to the shot noise). Some constant is indeed seen in N-body sims and can be of either sign depending on the mass bin etc.. Anyway the amplitude is small and I am not sure this approach will do a good job at predicting it. If we were more careful we would probably define bias once any constant is subtracted from $P_{hh}$.}

Finally, we look at terms of the form $\kappa_{n,m}^P \kappa_{n',m'}^Q$.  We may bound these contributions by using the above estimates with the convolution $\kappa_{n,m}^P \star \kappa_{n',m'}^Q = \int_\q \kappa_{m , n}^P (|\q|) \kappa_{n',m'}^Q (|\k-\q|)$.  For $n,m>1$, the convolution will be dominated by the IR, and we find 
\beq
\kappa_{n,m}^P \star \kappa_{n',m'}^Q \ \sim\ \kappa_{n,m}^P (k) \kappa_{n',m'}^Q(k)  \, \frac{\alpha^2(k) \ln (k L)}{P_{\m\m}(k)} \ .
\eeq
For $m = m' = 1$, the convolution is dominated by physics at the non-linear scale, so we may simply multiply (\ref{equ:koneX}) and/or (\ref{equ:koneone}) to find
\beq
\kappa_{n,1}^P \star \kappa_{n',1}^Q \ \sim \ \kappa_{n,1}^P (k) \kappa_{n',1}^Q(k) \, k_{\rm nl}^3 \ .
\eeq
As a result, convolutions of different cumulants will be suppressed by $ \fnl (1+\Pi)  \Delta_\Phi \ll 1$.

%For $P, Q > 1$, we may bound their contributions by products of the above estimates.  If $n,m >1$ and $n'=1$ and/or $m'=1$, this will be an overestimate.  \db{Please fix the next sentence:} By multiplying the above results, we are estimating a convolution integral the result using the peak values of the factors inside the integrand.  However, when $n,m >1$ and $n'=1$ and/or $m'=1$, these factors peak at large and small $k$, respectively, and therefore the integral is suppressed.  Nevertheless, the overestimate is proportional to $\Delta^{P+Q}$ \db{which $\Delta$?} so we may safely ignore it.

% I think this rather technical discussion of which cumulants to keep might be better suited for an appendix, since it distracts the reader from the main point which is the calculation of stochastic bias.}
%Using $\Delta_\m(k_{\rm eq}) < 1$ is the amplitude of the matter power spectrum at the scale where that entered the horizon at matter-radiation equality.  The suppression here is coming from that that the matter power spectrum has a peak at $k_{\rm eq}$, where the matter is still linear.  \dg{This might be too much detail.}

\subsection{$g_{\rm NL}$ Cosmology}
\label{app:convergence_gnl}

%\db{intro}
Similar arguments apply to the $\gnl$ model of \S\ref{sec:gnl}.

\vskip 4pt
\noindent
{\it Linear terms.}---First, we note that $\kappa_{n,m}=0$, unless  $n+m$ is even. Moreover, only for both $n$ and $m$ odd do we get a scale-dependent tree-level contribution to the cumulant. (In the main text, we discuss the important special case $\kappa_{2,2}$.) 
%\db{This is NOT the same as scale-dependence of the bias! e.g. constant cumulant gives scale-dependent bias. It is the relative scale-dependence to $\kappa_{1,1}$ that matters. We see this happening for $\kappa_{2,2}$ below!} \ssf{Again this is true, however $\kappa_{2,2}$ to tree level is just a small constant - which gives rise to a (small) $1/k$ bias through $P_{mm}$. We have a plot of the magnitude in our notes - but I'm not sure we should include it here. The following formulas don't apply to loops. I am not sure whether we should count $k_{2,2}$ as relevant, since so far people haven't seen stochasticity in $\gnl$ - once we decide if we want to count it or not, then we can improve this section.}
Schematically, we can write $\kappa_{n,m} \sim \gnl^{\beta} \ \Delta_{\Phi}^{\gamma}$. %\db{normalization?} 
%\ssf{just want to say that we only keep those cumulant that have the power of $\Delta_{\Phi}$ bigger than that of $\gnl$ (simply because $\gnl$ is potentially big...)  }.
%As a rule of thumb, we can neglect terms with $\gamma > \beta$, since only $\gnl \gtrsim 10^5 \sim \Delta_\Phi^{-1}$ is realistically observable. % \db{citation}. 
%\kms{I don't see why we need the preceding 2 sentences}
At tree level, we then find 
\bea \kappa_{1,m} & \sim & \gnl^{\frac{m-1}{2}} \ \Delta_{\Phi}^{m-2} \cdot \frac{P_{\rm mm}(k)}{\alpha(k)}  \ \ \ \ \ \ \ \ \quad \ \ \, \mathrm{for \ \ } m \mathrm{\ odd}\ , \label{eq:gnl_schematic2} \\
 \kappa_{n,m} & \sim & \gnl^{\frac{n+m}{2}-1} \ \Delta_{\Phi}^{m+n-4} \cdot \frac{P_{\rm mm}(k)}{\alpha^2(k)}  \ \ \ \ \ \ \ \mathrm{for \ \ } n,m \mathrm{\ odd \ \, and \, } >1 \label{eq:gnl_schematic}\ .
\eea
Since current observational constraints imply $|\gnl \Delta_\Phi^2 | \ll 1$,
the only tree-level terms that we need to keep are $\kappa_{1,1}$, $\kappa_{1,3} = \kappa_{3,1}$ and $\kappa_{3,3}$. 
As we discuss in the main text, there is also an interesting loop contribution to $\kappa_{2,2}$.

\vskip 4pt
\noindent
{\it Non-linear terms.}---As in the $\tnl$ model, products of cumulants of the form $\kappa_{n,m}^P$ will be suppressed due to the near-Gaussianity of the perturbations.  The contributions of higher powers of $\kappa_{n,m}$ is nearly identical in both cases:

\begin{itemize}
\item When $n > 1$ and $m>1$, %{and} odd, 
we take powers of the contributions in (\ref{eq:gnl_schematic}), to find    
\begin{align}
\kappa_{n,m}^P  &\ \sim\  \Big[ \, \gnl^{\frac{n+m}{2}-1}  \Delta_{\Phi}^{m+n-4}\, \Big]^P \cdot \frac{P_{\m \m}(k)}{\alpha^2(k)} \cdot \Delta_\Phi^{2(P-1)} \ln^{P-1}(k L)  \nonumber \\
&\ \sim\ \kappa_{n,m} \Big[\,\gnl  \Delta_{\Phi}^2 \, \Big]^{(P-1)(\frac{n+m-2}{2})}  \ln^{P-1}(kL) \ .
\end{align}
%where $L$ is an infrared cutoff.  
Clearly, if $\gnl \Delta_\Phi^2 \ll 1$, then the higher powers of $\kappa_{n,m}$ are suppressed (if we assume that the log is small).

\item When $n=1$ and $m>1$, %is odd different, 
we take higher powers of the results in (\ref{eq:gnl_schematic2}), to find for $P>1$
\begin{align}
\kappa^P_{1,m} &\ \sim\  \Big[ \, \gnl^{\frac{m-1}{2}}  \Delta_{\Phi}^{m-2} \Big]^P \Delta_\Phi^{P-1} P_{\m\m} (k_{\rm nl})\nonumber \\
&\ \sim\ \kappa_{1,m}\left[ \gnl \Delta_{\Phi}^{2} \right]^{(P-1)(\frac{m-1}{2} )} \alpha(k) \cdot \frac{P_{\m \m}(k_{\rm nl})}{P_{\m \m}(k)}  \label{equ:kone}\ .
\end{align}
Again, we find that $P > 1$ contributions are suppressed by powers of $\gnl \Delta_\Phi^2 \ll 1$.  As in the $\tnl$ model, we find that $P=1$ has a different scaling with $k$ from $P>1$.
\end{itemize}
It should be clear that other cumulants behave in the same way as in the $\tnl$ model and will be suppressed by factors of $\gnl \Delta_\Phi^2$.

\newpage
 \begingroup\raggedright\endgroup

\end{document}